\documentclass[twocolumn]{aastex62}

\usepackage{amsmath}

\newcommand\msun{M_{\odot}}
\newcommand\mhi{M_{HI}}
\newcommand\LG{L_{g}}
\newcommand\lsun{L_{\odot}}
\newcommand\vsys{V_{sys}}
\newcommand\wfty{W_{50}}
\newcommand\wftyc{W_{50,c}}
\newcommand{\kms}{{\ensuremath{\mathrm{km\,s^{-1}}}}}
\newcommand\mhilim{M^{lim}_{HI}}
\newcommand\mbary{M_{bary}}

\submitjournal{ApJ}

\shorttitle{\ion{H}{1} in UDGs}
\shortauthors{Karunakaran et al.}

\begin{document}

\title{Systematically Measuring Ultra Diffuse Galaxies in \ion{H}{1}: Results from the Pilot Survey}

\correspondingauthor{Ananthan Karunakaran}
\email{a.karunakaran@queensu.ca}

\author[0000-0001-8855-3635]{Ananthan Karunakaran}
\affiliation{Department of Physics, Engineering Physics and Astronomy, Queen’s University, Kingston, ON K7L 3N6, Canada}
\author[0000-0002-0956-7949]{Kristine Spekkens}
\affiliation{Department of Physics and Space Science, Royal Military College of Canada P.O. Box 17000, Station Forces Kingston, ON K7K 7B4, Canada}
\affiliation{Department of Physics, Engineering Physics and Astronomy, Queen’s University, Kingston, ON K7L 3N6, Canada}

\author[0000-0002-5177-727X]{Dennis Zaritsky}
\affiliation{Steward Observatory, University of Arizona, 933 North Cherry Avenue, Rm. N204, Tucson, AZ 85721-0065, USA}
\author[0000-0001-7618-8212]{Richard L. Donnerstein}
\affiliation{Steward Observatory, University of Arizona, 933 North Cherry Avenue, Rm. N204, Tucson, AZ 85721-0065, USA}
\author[0000-0002-3767-9681]{Jennifer Kadowaki}
\affiliation{Steward Observatory, University of Arizona, 933 North Cherry Avenue, Rm. N204, Tucson, AZ 85721-0065, USA}
\author[0000-0002-4928-4003]{Arjun Dey}
\affiliation{NSF's National Optical-Infrared Astronomy Research Laboratory, 950 N. Cherry Ave., Tucson, AZ 85719, USA}

\begin{abstract}We present neutral hydrogen (\ion{H}{1}) observations using the Robert C.\ Byrd Green Bank Telescope (GBT) of 70 optically-detected UDG candidates in the Coma region from the Systematically Measuring Ultra-Diffuse Galaxies survey (SMUDGes).\ We detect \ion{H}{1} in 18 targets, confirming 9 to be gas-rich UDGs and the remainder to be foreground dwarfs.\ None of our \ion{H}{1}-detected UDGs are Coma Cluster members and all but one are in low-density environments.\ The \ion{H}{1}-detected UDGs are bluer and have more irregular morphologies than the redder, smoother candidates not detected in \ion{H}{1}, with the combination of optical color and morphology being a better predictor of gas richness than either parameter alone.\ There is little visual difference between the gas-rich UDGs and the foreground dwarfs in the SMUDGes imaging, and distances are needed to distinguish between them.\ We find that the gas richnesses of our HI-confirmed UDGs and those from other samples scale with their effective radii in two stellar mass bins, possibly providing clues to their formation.\ We attempt to place our UDGs on the baryonic Tully-Fisher relation (BTFR) using optical ellipticities and turbulence-corrected \ion{H}{1} linewidths to estimate rotation velocities, but the potential systematics associated with fitting smooth $\mathrm{S\acute{e}rsic}$ profiles to clumpy, low-inclination low surface brightness disks precludes a meaningful analysis of potential BTFR offsets.\ These observations are a pilot for a large campaign now underway at the GBT to use the \ion{H}{1} properties of gas-rich UDGs to quantitatively constrain how these galaxies form and evolve.\ \end{abstract}


\section{Introduction} \label{sec:intro}
The study of the low surface brightness (LSB) galaxy population has been reinvigorated as a result of improvements in astronomical instrumentation \citep[e.g.,][]{DragonflyArrayOriginal,HSC-SurveyOverview} and data reduction methods \citep[e.g.,][]{IACStripe82,2016Trujillo}, as well as the use of novel image searching algorithms \citep[e.g.,][]{Paulspaper,LeoILeoTriplet,2018Proledetectionmethod,SMUDGes,2020Carlsten}.\ Among this recent surge of LSB detections are populations of extended red LSB galaxies akin to those discovered in early LSB studies \citep[e.g.,][see also \citealt{Conselice}]{SandageUDG,1988Impey,1991Bothun}.\ 

In their survey of the Coma Cluster, \citet{UDGsvandokkum} presented the first significant sample of these extended LSBs, dubbing them ultra diffuse galaxies (UDGs) and proposing size $(R_{eff} > 1.5 \,\mathrm{kpc})$ and surface brightness $(\mu_{0,g}\gtrsim24\,\mathrm{mag/arcsec^2})$ criteria that have since been widely adopted to define them.\ To date, over 1000 UDG candidates have been discovered in subsequent searches of the Coma Cluster \citep[e.g.,][]{UDGs-1Koda,UDGs-4Yagi,SMUDGes} and several other clusters \citep[e.g.,][]{2015MihosVirgoUDGs,2016BeasleyVirgoUDG,UDGs-8Shi,2017VenholaFornaxUDGs,2019ManceraClusters,2020LeeDistantClusterUDGs}, as well as a growing number in lower density environments \citep[e.g.,][]{UDGs-3MartinezDelgado,Secco-dI-1and2,UDGs-7RomanTrujillo,UDGs-6Leisman,UDGs-5Trujillo,PaulTidalUDG,2019Roman,2020SPLUS}.\ 

Across these environments there exists a large diversity in the physical properties of UDGs, similar to that seen in the high surface brightness galaxy population.\ Most UDGs seem to be embedded in dwarf galaxy-mass dark matter halos \citep[e.g.,][]{2016BeasleyGC-UDGsdwarfs,2018AmoriscoGlobCluster,internaldynamicsUDGs,2019Prolehalomass}, although there is evidence that at least some are in more massive halos (\citealt[][]{VelocityDispersionvandokkum,2017Zaritsky,2018Lim,2020Forbes-twotypesofUDGs}, although see \citealt{2020saifollahi}).\ While UDGs found in clusters tend to be red (i.e., quiescent) and smooth, those in lower density environments are bluer (i.e., star forming) and have more irregular morphologies \citep{UDGs-7RomanTrujillo,2019prole}.\ Some UDGs exhibit extreme properties that pose challenges to proposed galaxy formation mechanisms, such as high dark matter fractions \citep{VelocityDispersionvandokkum,2016BeasleyVirgoUDG}, dark matter deficiencies (\citealt{lackingdm,lackingdm3}; although see \citealt{lackingdm2}), and offsets from established galaxy scaling relations such as the baryonic Tully-Fisher relation \citep{2019HUDsBTFR,2020ManceraPina}.\ 

Proposed UDG formation mechanisms generally fall into two categories: internally and externally-driven physics.\ Isolated (i.e., field) UDGs may be formed through multiple internal mechanisms.\ For example, \citet{UDG-Formation1} suggest that UDGs formed in dwarf dark matter halos with elevated angular momenta, naturally explaining their extended sizes.\ Alternatively, using the NIHAO (Numerical Investigation of a Hundred Astrophysical Objects, \citealt{nihao}) suite of simulations, \citet{UDG-Formation2} show that UDG-like objects can form through bursty star-formation early in their evolution resulting in a more extended, diffuse matter distribution.\ The red, smooth UDGs observed in groups and clusters may represent the population of field UDGs that formed through the aforementioned mechanisms and were subsequently quenched via ram-pressure and/or tidal effects \citep{YozinUDGs,2019Liao,2019Jiang,tidalUDGs}.\ However, some may form initially as typical dwarf galaxies that are tidally disturbed after in-fall into a cluster or by a massive companion \citep{PaulTidalUDG,2020SalesIllustrisUDGs}.

In order to constrain which of these proposed formation mechanisms explains the origin of the detected UDGs, larger samples of UDGs with distance measurements are required, particularly in the field where inferring distances by projected separation from clusters or groups is not possible.\ While some optical distances to UDGs have been obtained \citep{DF44confirmationvandokkum,Secco-dI-1and2,Kadowaki2017,2018Alabi-spectro,ComaopticalfollowupI,ComaopticalfollowupII,internaldynamicsUDGs,ChemicalabundanceUDG}, sample sizes are limited due the large spectroscopic integration times required at low surface brightnesses.\ 

By contrast, the neutral hydrogen (\ion{H}{1}) in gas-rich UDGs can not only provide a distance measure but also help distinguish among formation mechanisms.\ The \ion{H}{1} redshift provides kinematic distances for candidates that can distinguish foreground dwarfs $(R_{eff} < 1.5\,\mathrm{kpc})$ from true UDGs $(R_{eff} > 1.5\,\mathrm{kpc})$, and linewidths reflect their internal dynamics, and the \ion{H}{1} flux provides the gas mass.\ \ion{H}{1} follow-up observations of optically-detected UDG candidates have been demonstrated to be feasible with single-dish radio telescopes \citep{IsolatedUDGs-HI,HCGUDGs} and searches through extant blind \ion{H}{1} survey detections for diffuse stellar counterparts have also been fruitful \citep{UDGs-6Leisman}.

The Systematically Measuring Ultra-Diffuse Galaxies \citep[SMUDGes;][hereafter Z19]{SMUDGes} survey is uniquely positioned to produce samples of UDG candidates for \ion{H}{1} follow-up observations across a range of environments, as the combination of depth and coverage of the DECaLS data used to detect UDG candidates are unmatched.\ The SMUDGes pilot survey searched publicly available DECaLS data (one of three DESI pre-imaging Legacy surveys, see \citealt{Dey2019desi} for details) for large $(r_{eff} > 5.3\arcsec = 2.5 \, \mathrm{kpc\,at\,}D_{Coma}\sim100\mathrm{Mpc})$ UDG candidates in a $290\, \mathrm{deg^2}$ region centered on the Coma Cluster.\ The 275 UDG candidates resulting from that search (Z19) as well as subsequent SMUDGes detections provide ample targets to pilot a large follow-up campaign.\ 

In this paper, we present pilot \ion{H}{1} observations along the lines-of-sight to 70 SMUDGes UDG candidates, which represent the first phase of a large \ion{H}{1} follow-up campaign using the Robert C.\ Byrd Green Bank Telescope (GBT).\ We aim to obtain redshift measurements to UDG candidates and characterize the gas properties of confirmed UDGs to constrain their formation mechanisms.\ These observations, which are part of a much larger GBT program that is currently underway, represent the largest \ion{H}{1} follow-up campaign of optically-selected UDG candidates ever reported.

The structure of this paper is as follows.\ In Section \ref{sec:sample}, we describe our \ion{H}{1} target selection.\ We outline our observations and data reduction procedure in Section \ref{sec:Obsanddata}.\ In Section \ref{sec:results}, we present the properties of our \ion{H}{1} detections and non-detections.\ In Section \ref{sec:Discussion}, we discuss the environmental and morphological properties of \ion{H}{1} detections and non-detections, place initial constraints on UDG formation mechanisms, and discuss our UDGs in the context of the baryonic Tully-Fisher relation.\ We conclude and outline future work in Section \ref{sec:Conclusion}.\ Throughout this work we use $D_{Coma} = 100\,\mathrm{Mpc}$, $H_0 = 70 \kms \mathrm{Mpc^{-1}}$, $\Omega_\Lambda = 0.7$, and $ \Omega_m = 0.3.$

\section{Sample Selection} \label{sec:sample}
We select \ion{H}{1} follow-up targets from the SMUDGes pilot sample (Z19) and subsequent searches of the DECaLS data.\ Focused on the $290\, \mathrm{deg^2}$ region centered on the Coma Cluster, the SMUDGes pilot survey employed a semi-automated UDG candidate identification procedure, described in detail in Z19.\ Briefly, the DECaLS observations were preprocessed to remove any defects, and then foreground or background sources significantly brighter than UDGs were replaced with background noise.\ Next, these processed images were spatially filtered to various scales using wavelet transforms, and diffuse objects are identified using SEP \citep{2016SEP,1996SourceExtractor}.\ In order to compare results with other studies \citep[e.g.,][]{UDGsvandokkum,UDGs-4Yagi}, their photometric properties were then modeled as exponential profiles using GALFIT \citep{Galfit} and only objects with $r_{eff} > 5.3\arcsec (R_{eff}=2.5 \, \mathrm{kpc\,at\,}D_{Coma}\footnote{We use $r_{eff}$ for angular sizes and $R_{eff}$ for physical sizes throughout this paper.})$ and $\mu_{0,g}> 24\, \mathrm{mag\,arcsec^{-2}}$ were kept.\ The remaining objects were examined by eye, and 275 classified as bona-fide UDG candidates.\

We select 34 of the 275 SMUDGes UDG candidates with $m_g \lesssim 19.5\,\mathrm{mag}$  to follow up in \ion{H}{1} (listed as Z19 in column 14 of Table \ref{table:maintable}).\ This magnitude limit combined with gas richness scaling relations for local dwarfs \citep[e.g.,][]{Bradford2015MAIN} implies that integration times of no more than a few hours are required to follow up each source (see Section \ref{sec:Obsanddata}).\ A subsequent optical search within the same region using an improved SMUDGes pipeline (Zaritsky et al., in prep) detected an additional 36 UDG candidates that satisfied the above magnitude limit and we include them in our \ion{H}{1} follow-up sample as well (listed as K20 in column 14 of Table \ref{table:maintable}).\ The DECaLS imaging for all targets was subsequently modeled as a $\mathrm{S\acute{e}rsic}$ profile with a variable $\mathrm{S\acute{e}rsic}$ index using GALFIT and the resulting parameters are listed in columns (4)-(11) of Table \ref{table:maintable}.\ The parameter uncertainties are the GALFIT values which are derived using Poisson pixel noise; a more comprehensive error estimation method for SMUDGes photometery is being developed using simulated UDG recovery for use in the full survey \citep{SMUDGes}.\  

The optical properties of the 36 previously unpublished UDG candidates are largely consistent with the sample selected from Z19, although there are a few candidates with smaller sizes $(r_{eff} > 4.7\arcsec)$ and higher surface brightnesses $(\mu_{0,g} > 23.7\,\mathrm{mag\,arcsec^{-2}})$.\ Some of these candidates have $m_g > 19.5\,\mathrm{mag}$ (our \ion{H}{1} follow-up criterion) because initial estimates were used during the target selection.\ There is some overlap of the UDG candidates in SMUDGes with other UDG samples (see Z19).\ Of the 36 UDG candidates we present here, 6 have been either presented in other work and/or previously detected in \ion{H}{1}.\ We include references for these objects in column 14 of Table \ref{table:maintable}.

\begin{figure*}[htb!]
\includegraphics[width=18cm]{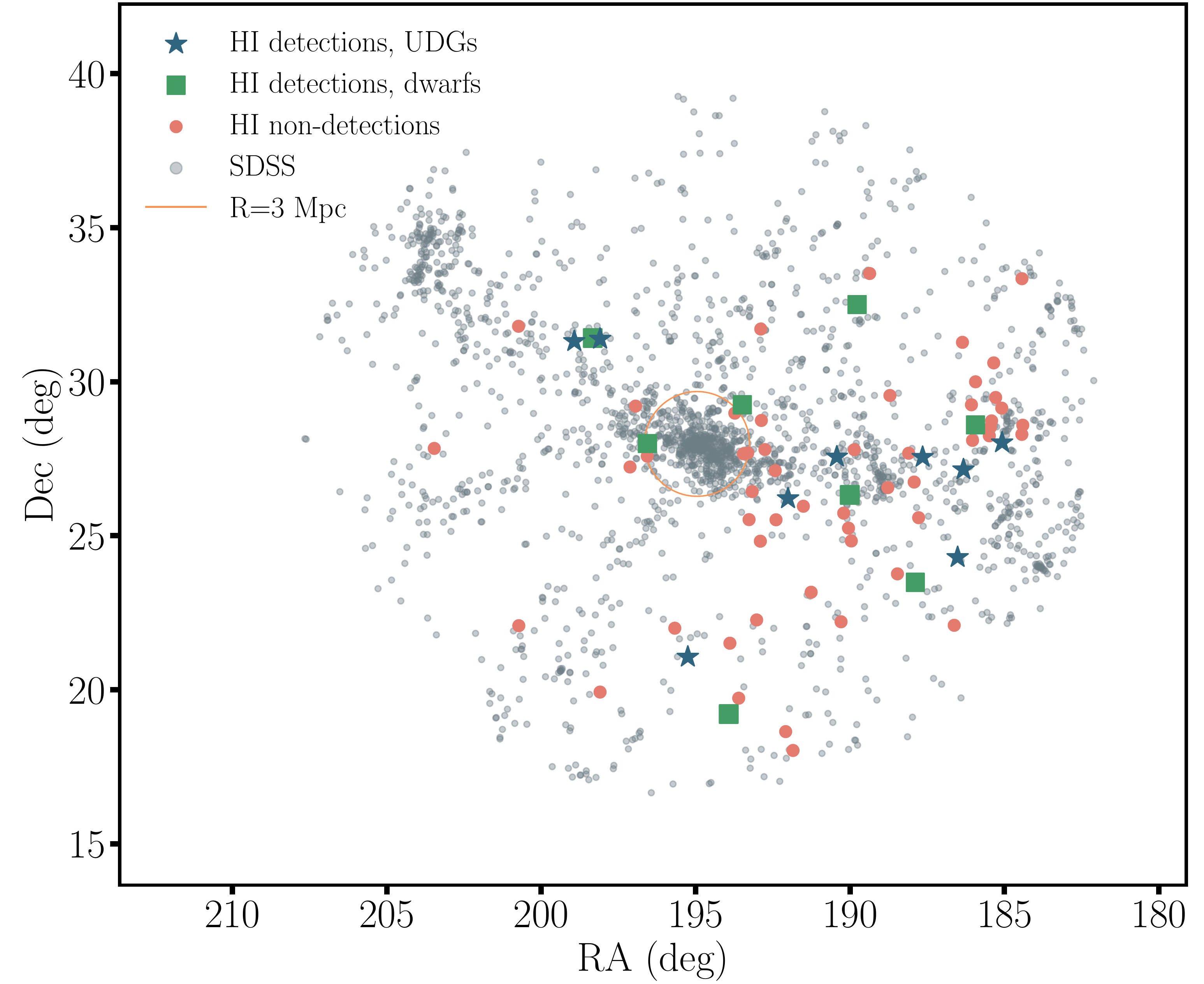}
\caption{Projected sky distribution of our UDG candidate \ion{H}{1} follow-up sample in the Coma Cluster region (colored points), with galaxies from SDSS DR15 with $5000 \kms < cz < 9000 \kms$ plotted as small grey circles.\ Our sample is subdivided into \ion{H}{1} detections of UDGs (blue stars), \ion{H}{1} detections of foreground dwarf galaxies (green squares), and \ion{H}{1} non-detections (red circles).\ The orange open circle is centered on the Coma Cluster \citep[$\alpha=12^h59^m48.7^s$; $\delta=27^{\circ} 58'50''$,][]{Kadowaki2017} and has a radius of $\sim3$ Mpc that represents the virial radius of the Coma Cluster \citep{comavirradius}.\ }
\label{fig:skydist}
\end{figure*}

In total, our \ion{H}{1} follow-up sample consists of 69 UDG candidates in the Coma Cluster region and 1 outside of it\footnote{The exception is SMDG1103517+284118 which falls outside the Coma Cluster region and was nonetheless included as a target of interest}.\ Their projected spatial distribution relative to galaxies from the SDSS DR15 \citep{SDSSDR15} with $5000 \kms < cz < 9000 \kms$ is shown in Figure \ref{fig:skydist}.\ 

We do not select on color in this work despite its accuracy for predicting gas richness in the high surface brightness galaxy population \citep[e.g.,][]{GASS,Brown2015}.\ Instead, we investigate the relationship between color and gas-richness in Sections \ref{subsec:HInondetect} and \ref{sec:Discussion}.\

\section{Observations and Data Reduction} \label{sec:Obsanddata}
We performed 88 hours of position-switched \ion{H}{1} observations between 2018 February and 2018 August using the GBT along the lines of sight (LOS) to the 70 UDG candidates in Table \ref{table:maintable} (program AGBT18B-239).\ 9 objects were observed with an offset between the optical centroid and the LOS in order to minimize contamination from known nearby objects (see Section \ref{subsec:detections}).\ These objects are indicated with an asterisk next to their RA in Table \ref{table:maintable}.

Our observational setup and data reduction procedure are similar to those used in \citet{KarunakaranM101}, which we briefly outline here.\ We used the L-band receiver and the Versatile GBT Astronomical Spectrometer (VEGAS) with a spectral resolution of 3.1 kHz and a wide bandpass of 100 MHz which allows for the detection of \ion{H}{1} emission lines out to $V_{Helio} \sim 14000 \, \kms$.\ We estimate the integration times for our targets using $m_g$ to reach a gas richness of $\frac{\mhi}{\LG} \sim$ 1 $\frac{\msun}{\lsun}$ with $S/N = 5$ in a single $50 \, \kms$ channel.\ Gas richness is a distance-independent quantity since both $\mhi$ and $\LG$ scale with distance squared.\ Therefore, a single spectrum allows us to search for an \ion{H}{1} reservoir in our targets anywhere within the wide bandpass.\

The data were reduced using the standard GBTIDL\footnote{http://gbtidl.nrao.edu/} procedure $getps$.\ We remove narrow-band and broadband radio frequency interference before smoothing our spectra to our desired resolutions, following the same procedure as \citet{KarunakaranM101}.\ Furthermore, we scale the fluxes in our final spectra up by 20\% to account for the systematic offset in the GBT noise diode calibration values reported by \citet[][]{gbtcal}.\ The RMS noise, $\sigma_{50}$, for each spectrum at $\Delta V = 50\,\kms$ resolution is given in column 13 of Table \ref{table:maintable}.\

We examined the calibrated, RFI-excised spectra by-eye after smoothing to multiple resolutions from $5-50 \, \kms$ for statistically significant emission.\ We detect \ion{H}{1} emission along the LOS to 18 UDG candidates (column 15 of Table \ref{table:maintable}).\  We show their spectra in Figure \ref{fig:detectspectra} at $\Delta V$ given in Tables \ref{table:detectiontable} and \ref{table:dwarftable}, which also lists other properties we have derived from these \ion{H}{1} detections.\ We find no significant \ion{H}{1} emission associated with the 52 remaining targets and place stringent $5\sigma$ the upper limits on \ion{H}{1} mass, ${\mhilim}$, and gas richness, ${\mhilim}/{\LG}$, which are listed in Table \ref{table:upperlimits}.

\section{Results} \label{sec:results}
\subsection{Properties of \ion{H}{1} Detections} \label{subsec:detections}
\begin{figure*}[htb!]
\includegraphics[width=18cm]{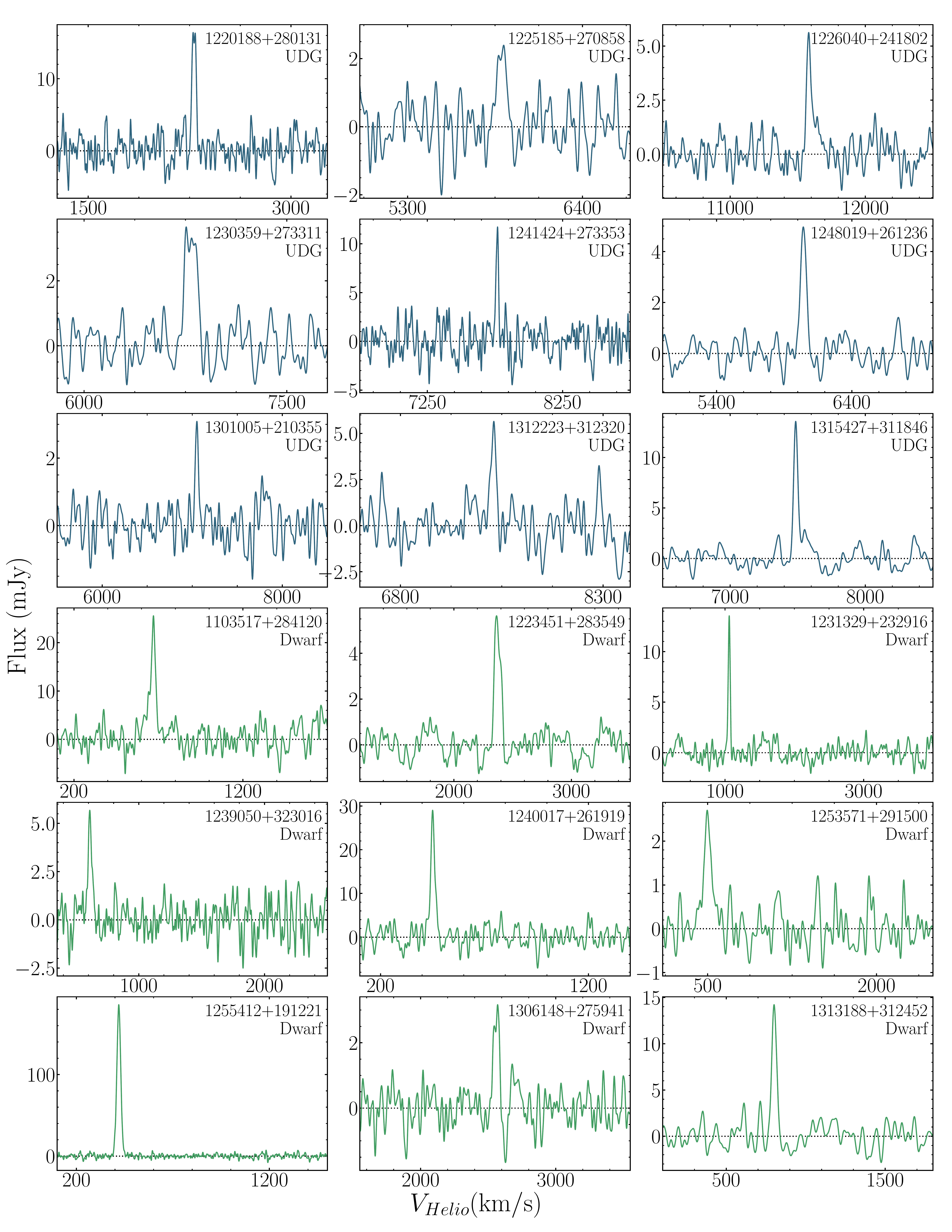}
\caption{\ion{H}{1} detections along the LOS to UDG candidates in our sample.\ The first 9 panels show targets that satisfy the UDG size criterion of $R_{eff} > 1.5 \mathrm{kpc}$ given their redshifts (confirming them as UDGs), while the last 9 panels show targets which do not (confirming them as foreground dwarfs).\ Target names and classification (UDG or Dwarf) are in the top-right corner of each panel.\ The black dotted line in each panel represents 0 mJy.\  The spectral resolutions $\Delta V$ of the plotted spectra and the derived properties of the \ion{H}{1} detections are in Tables \ref{table:detectiontable} and \ref{table:dwarftable} for UDGs and foreground dwarfs, respectively.}
\label{fig:detectspectra}
\end{figure*}

We detect \ion{H}{1} along the LOS to 18 UDG candidates, and their spectra are shown in Figure \ref{fig:detectspectra}.\ At our observing frequency of $\sim$ 1.4 GHz, the GBT beam (FWHM $\sim \mathrm{9.1'}$) response is well understood down to $\approx -30 \mathrm{dB}$ \citep[e.g.,][]{GBTbeam}.\ We therefore search through NED\footnote{The NASA/IPAC Extragalactic Database (NED) is operated by the Jet Propulsion Laboratory, California Institute of Technology, under contract with the National Aeronautics and Space Administration.} and the DESI Legacy Imaging Survey Sky Viewer\footnote{http://legacysurvey.org/viewer} for objects within $30'$ of the LOS that may present themselves as gas-rich interlopers in our spectra for all of our targets.\ We find no such interlopers for any of our \ion{H}{1} detections, and conclude they are the \ion{H}{1} counterparts to the corresponding UDG candidates.\ 

We derive distance-independent quantities from the spectra (systemic velocity, $\vsys$, and velocity width, $\wfty$) as described in \citet{KarunakaranM101} and briefly outline the method here.\ Using a first-order polynomial fit to each edge of the \ion{H}{1} profile between 15\% and 85\% of the peak flux value, we find the velocities corresponding to the 50\% flux value.\ Their mean corresponds to $\vsys$ (column 4 of Tables \ref{table:detectiontable} and \ref{table:dwarftable}) and the difference corresponds to $\wfty$.\ The latter is corrected for instrumental and cosmological redshift broadening following \citet{2005ApJS..160..149S} to produce $\wftyc$ (column 5 of Tables \ref{table:detectiontable} and \ref{table:dwarftable}).\ We assume an uncertainty of 50\% for the instrumental broadening correction, which dominates the uncertainties on $\vsys$ and $\wftyc$.\ We note that we are conducting signal recovery simulations, similar to \citet{2005ApJS..160..149S} but tailored to UDG-like \ion{H}{1} profile shapes, to more accurately understand how instrumental effects at the GBT affect our \ion{H}{1} detections in the full survey.\

Distances are required to confirm candidates as true UDGs.\ Using $\vsys$ and the Hubble-Lema\^{i}tre Law, we estimate kinematic distances for all of our \ion{H}{1} detections and adopt a distance uncertainty of 5 Mpc \citep{UDGs-6Leisman, HCGUDGs}.\ Interestingly, detections are almost equally split between foreground $(D_{HI} < 40 \,\mathrm{Mpc})$ and background objects $(D_{HI} > 80 \,\mathrm{Mpc})$: this emphasizes how gas-rich diffuse objects at different distances can look similar on the sky (see also Figures \ref{fig:arcomp} and \ref{fig:dwcomp}), an issue that we discuss further in Section \ref{subsec:gasrich}.\ Based on these distances and the angular sizes listed in Table \ref{table:maintable}, we confirm 9 new UDGs with $R_{eff} > 1.5\,\mathrm{kpc}$ and $\mu_{0,g}\gtrsim 24 \,\mathrm{mag\,arcsec^{-2}}$, and give their \ion{H}{1} properties in Table \ref{table:detectiontable}.\ The remaining detections are dwarfs in the foreground of Coma; their derived \ion{H}{1} properties are in Table \ref{table:dwarftable}.\ To the best of our knowledge, the UDGs in Table \ref{table:detectiontable} represent the largest sample of optically-selected UDGs with follow-up \ion{H}{1} detections reported so far.

\begin{figure*}[htb!]
\includegraphics[width=18cm]{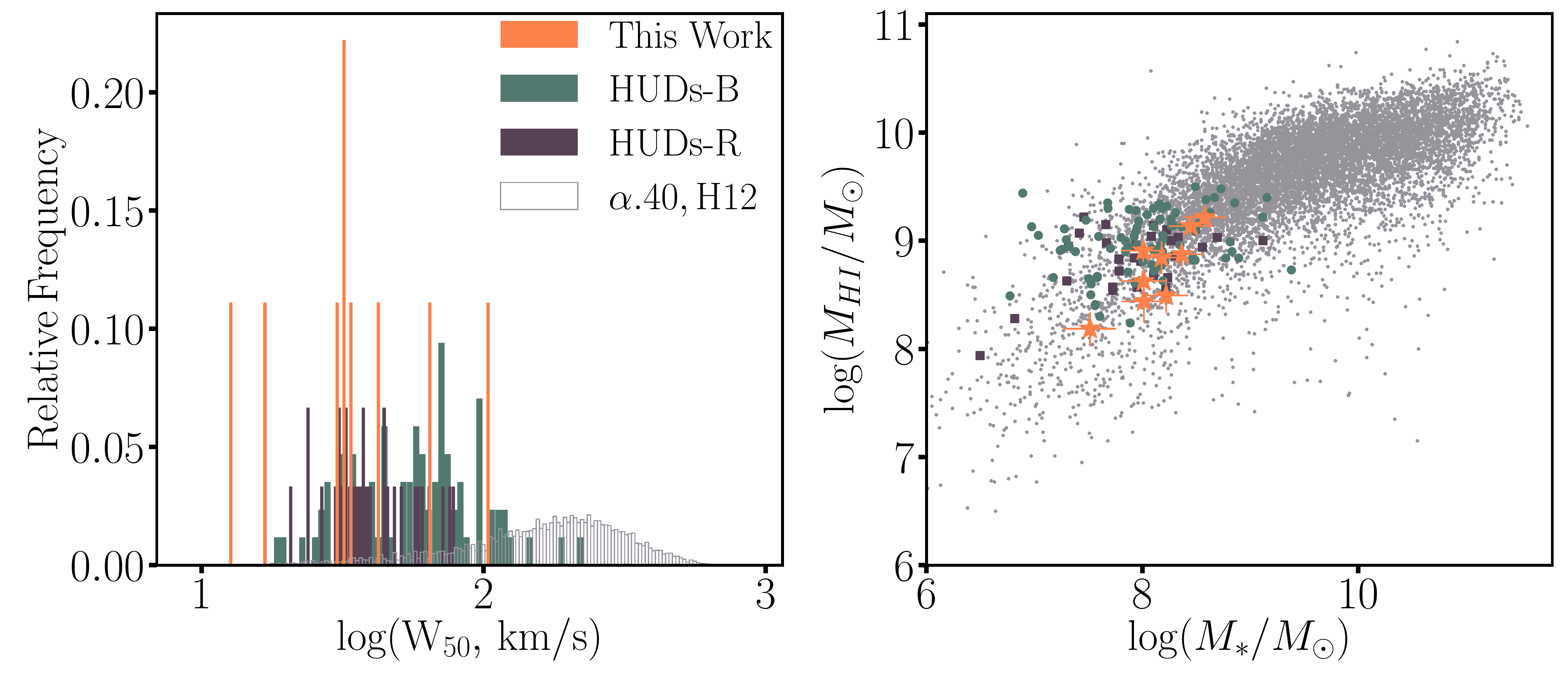}
\caption{Comparison of derived properties between our \ion{H}{1}-confirmed UDGs and other similar samples.\ $Left:$ Distribution of $\wftyc$ for UDGs in our sample (orange), the HUDs-B and -R samples (purple and green, respectively), and galaxies from the $\alpha.40$ catalog with SDSS and GALEX coverage (grey).\ $Right:$ $\mhi - M_{*}$ relation for the same samples as in the left panel.\ }
\label{fig:hipropcomp}
\end{figure*}

In the left panel of Figure \ref{fig:hipropcomp}, we compare the distribution of $\wftyc$ for our \ion{H}{1}-confirmed UDGs (orange) to those of the \ion{H}{1}-bearing ultra-diffuse sources (HUDs) samples, HUDs-B (green) and HUDs-R (purple), from \citet{UDGs-6Leisman}.\ The HUDs-B and -R samples are distinguished by their ``broad" and ``restrictive" optical selection criteria (see \citealt{UDGs-6Leisman} for more details) with the latter sample using the same criteria used in this work.\ We also include galaxies in the ALFALFA $\alpha.40$ catalog \citep{ALFALFAFourtyPercent}.\ Our \ion{H}{1}-confirmed UDGs span a broad range in $\wftyc$ and are generally more consistent with the HUDs samples than galaxies in ALFALFA.\ During our literature search for possible \ion{H}{1} interlopers we discovered that 5 of our 18 \ion{H}{1} detections were previously reported as gas-rich objects.\ Of these previously detected objects, 1 is a UDG (SMDG1220188+280132) that has been detected by ALFALFA.\ It was not included in the HUDs sample, likely due to their distance criteria \citep{UDGs-6Leisman}.\ Therefore, we present SMDG1220188+280132 as a UDG here for the first time.\ 

We calculate the \ion{H}{1} flux, $S_{HI}={\int}S{\delta V}$, by integrating over the line profile, where uncertainties stem mainly from the noise statistics of the profile \citep{2005ApJS..160..149S} and a 2\% noise diode uncertainty \citep{1997AJ....113.1638V}.\ We use these fluxes and our kinematic distances to determine \ion{H}{1} masses, $\mhi$, using the standard equation for an optically thin gas \citep{1984AJ.....89..758H}: \begin{equation} \label{eqn:himass} M_{HI}=2.356\times 10^{5}(D_{HI})^{2}S_{HI} \, \mathrm{\msun},\end{equation} where the distance, $D_{HI}$, is in Mpc and $S_{HI}$ is in Jy $\kms$.\ \ion{H}{1} masses are listed in column 8 of Tables \ref{table:detectiontable} and \ref{table:dwarftable}.\ Uncertainties are determined following the methods of \cite{2005ApJS..160..149S} to which distance uncertainties are added in quadrature.\ 

 We calculate stellar masses, $M_{*}$, for detections (column 9 of Tables \ref{table:detectiontable} and \ref{table:dwarftable}) using $m_{g}$ and $g-r$ from Table \ref{table:maintable} in the relations of \citet[][]{2017zhang} and assuming $D_{HI}$, propagating photometric and distance uncertainties along with those reported on the relations.\ Finally, we estimate baryonic masses as $\mbary=1.33\mhi+M_{*}$ (column 10 of Tables \ref{table:detectiontable} and \ref{table:dwarftable}).\ 
 
 In the right panel of Figure \ref{fig:hipropcomp}, we show the \ion{H}{1}-confirmed UDGs in the $\mhi-M_{*}$ plane, along with the HUDs samples (-B: green circles and -R: purple squares) and galaxies from the $\alpha.40$ catalog with SDSS and GALEX coverage from \citet[][grey circles]{huang2012galaxypop}.\ We find that our UDGs are broadly consistent with both the HUDs and $\alpha.40$ samples, although Figure \ref{fig:hipropcomp} illustrates how the HUDs sample as a whole may be more gas-rich \citep[mean $\mhi/M_{*} \sim 15$,][]{UDGs-6Leisman} compared to ours $(\mathrm{mean\,}\mhi/M_{*} \sim 5)$.\ This may point to a difference in UDG samples drawn from \ion{H}{1} vs.\ optical searches, although their selection functions need to be understood before intrinsic population differences can be quantified.\ 

\begin{figure*}[htb!]
\includegraphics[width=18.5cm]{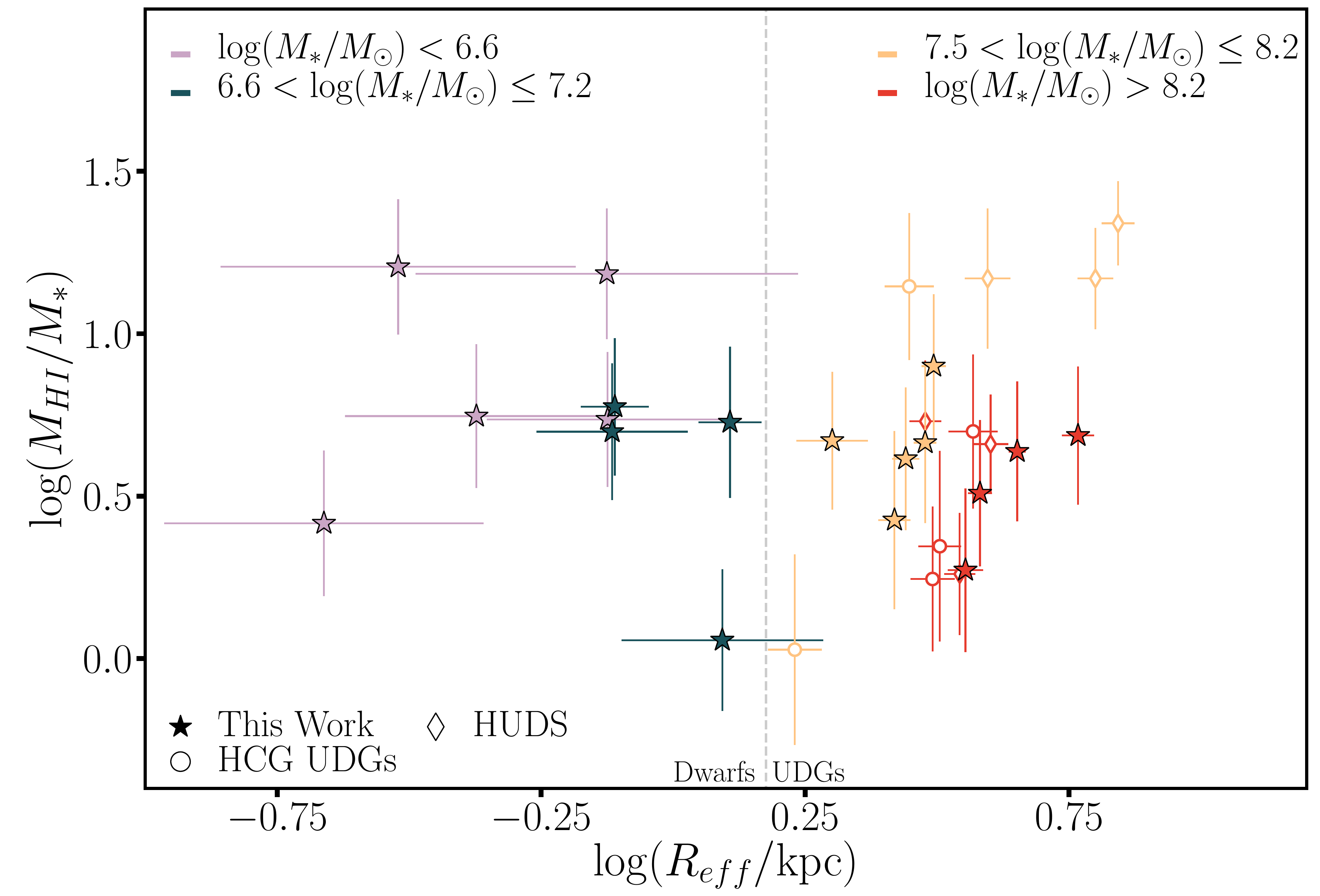}
\caption{Gas-richness as a function of size for our 9 \ion{H}{1}-confirmed UDGs and 9 foreground dwarfs as filled stars.\ The gray dashed line shows the $R_{eff}=1.5\, \mathrm{kpc}$ size boundary between dwarfs and UDGs.\ We also include the 5 UDGs around Hickson Compact groups from \citet[][open circles]{HCGUDGs} and 6 UDGs from \citet[][open diamonds]{2020ManceraPina}.\ The colors of the symbols represent the stellar mass bin of the objects.\ }
\label{fig:gasrichness-size}
\end{figure*}

In Figure \ref{fig:gasrichness-size}, we show the relationship between gas-richness, $\mhi/M_{*}$, and size, $R_{eff}$, for our \ion{H}{1}-confirmed UDGs and foreground dwarfs (filled stars).\ We also include the 6 UDGs from \citet[][]{2020ManceraPina} and 5 UDGs around Hickson Compact Groups from \citet[][]{HCGUDGs} (open symbols).\ For consistency across samples we have calculated $R_{eff}$ from the exponential scale-lengths reported by \citet{2020ManceraPina}.\ Because of the large systematic differences between different color-mass-to-light-ratio prescriptions \citep{2015Roediger}, we have also re-calculated stellar masses for the 5 UDGs followed up by \citet{HCGUDGs} using the photometry of \citet{UDGs-7RomanTrujillo} and the \citet{2017zhang} relations.\ We also propagated the 5 Mpc distance uncertainties into the errorbars on $R_{eff}$ for all samples.\ The colors of the symbols in Figure \ref{fig:gasrichness-size} represent the stellar masses of the object.\ There is some evidence that larger UDGs are more gas-rich within each stellar mass bin but little evidence for a similar trend among the foreground dwarfs; we discuss possible implications of this in Section \ref{subsec:formation}.\

\begin{figure*}[htb!]
\includegraphics[width=18.5cm]{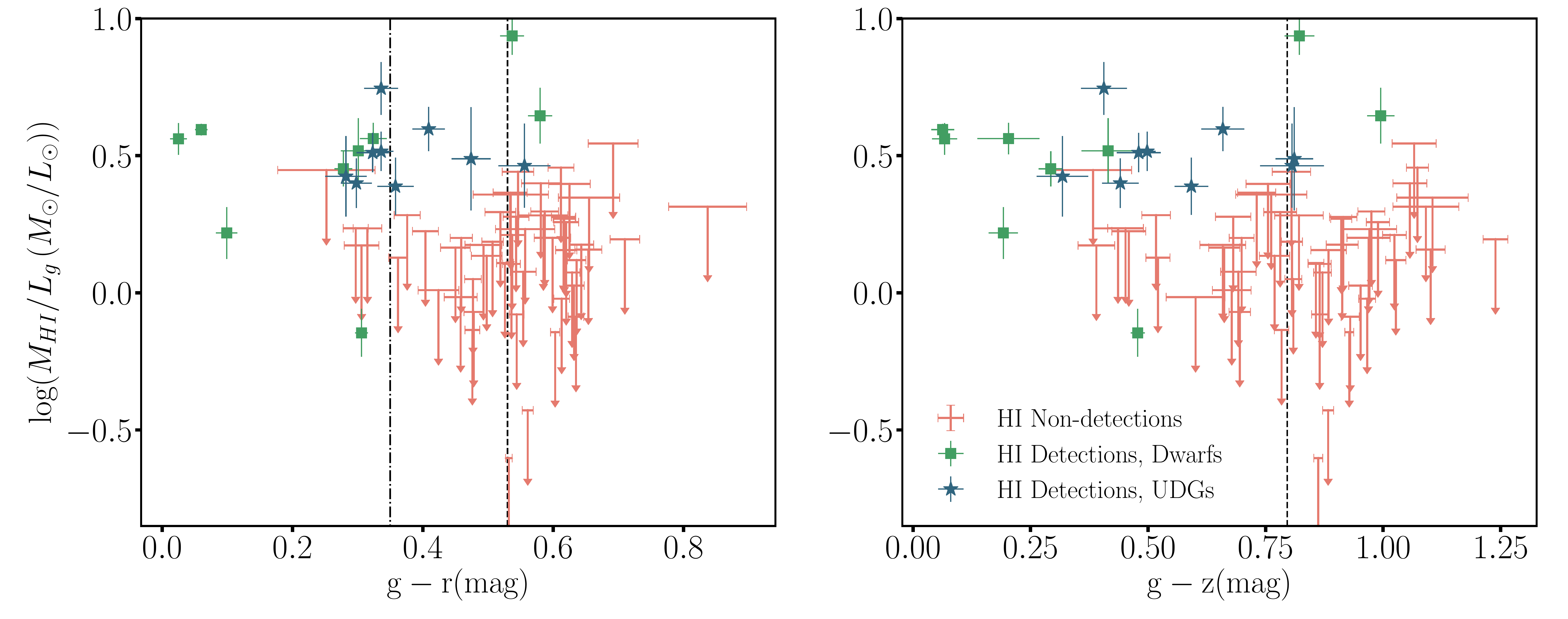}
\caption{$\mhi/\LG$ (blue stars and green squares for \ion{H}{1}-confirmed UDGs and foreground dwarfs, respectively) and $\mhilim/\LG$ (red downward arrows for non-detections) as a function of $g-r$ (left) and $g-z$ (right) color for our sample.\ The dashed vertical lines in each panel show the median $g-r=0.53$ and $g-z=0.79$ colors for our sample.\ For comparison, we show the median $g-r=0.35$ color of the HUDs sample \citep{UDGs-6Leisman} in the left panel with a vertical dash-dotted line.}
\label{fig:gasrichness-colour}
\end{figure*}

\subsection{\ion{H}{1} Non-detections} \label{subsec:HInondetect}
We find no statistically significant \ion{H}{1} signals along the LOS to 52/70 targeted UDG candidates that can be attributed to these objects.\ We smooth their spectra to $\Delta V = 50\, \kms$ and list their representative RMS noise, $\sigma_{50}$, in column 13 of Table \ref{table:maintable}.\ We modify equation \ref{eqn:himass} to place stringent, $5\sigma$ \ion{H}{1}-mass upper limits \begin{equation} \label{eqn:hilim}\mhilim=5.89\times 10^{7}(D_{lim})^{2}{\sigma_{50}}\, \mathrm{\msun},\end{equation} where $D_{lim}$ is the adopted distance in Mpc.\ In most cases we assume the Coma Cluster distance of $D_{lim}=D_{Coma}=100\,\mathrm{Mpc}$, aside from a few exceptions described below.\ We also set (distance-independent) upper limits on the ratio of \ion{H}{1}-mass to $g-$band luminosity, $\mhilim/\LG$, and list all of the calculated properties for non-detections in Table \ref{table:upperlimits}.

We briefly highlight a few of the \ion{H}{1} non-detections in our sample.\ SMDG1221577+281436 was reported as the marginal \ion{H}{1} detection of a nearby, gas-rich dwarf galaxy with $V_{Helio} = 450\pm8\,\kms$ in \citet[][d1221+2814]{huch2009}.\ However, when we smooth our spectra to match their velocity resolution we see no signal despite our deeper data.\ SMDG1253151+274115 (first reported as DF30 in \citealt{UDGsvandokkum}) and SMDG1251013+274753 were confirmed as UDGs via optical spectroscopy in \citet[][]{Kadowaki2017} and Kadowaki et al., in prep, with $V_{opt} = 7316 \pm 81 \, \kms$ and $V_{opt} = 6118 \pm 45 \, \kms$, respectively.\ The former was confirmed as a Coma Cluster member and we use $D_{lim}=100\,\mathrm{Mpc}$ to estimate \ion{H}{1} properties.\ The latter was confirmed to lie outside the Coma Cluster, and therefore we estimate its distance using $V_{opt}$ and the Hubble-Lema\^{i}tre Law to be $D_{lim}=87\,\mathrm{Mpc}$.\  In addition, Kadowaki et al.\ (in prep) find velocities for SMDG1217378+283519 $(V_{opt} = 493\pm69\,\kms)$ and SMDG1221086+292920 $(V_{opt} = 1024\pm66\,\kms)$ that place them well in the foreground of Coma, and we use the corresponding $D_{lim}$ to compute $\mhilim$.\ SMDG1302417+215954 (IC 4107) has previously been reported as an \ion{H}{1} non-detection \citep{1992schombert}.\ One SDSS spectrum of this object classifies it as a star\footnote{\url{http://cas.sdss.org/dr7/en/get/specById.asp?id=746142786461368320}} with $V_{opt} = 267\,\kms$  \citep{2014KimExVirgo}, while another classifies it as a QSO\footnote{\url{http://skyserver.sdss.org/dr12/en/get/SpecById.ashx?id=2983699425022470144}} with $V>100,000\,\kms$.\ We adopt $D_{lim}=3.8\,\mathrm{Mpc}$ using the lower SDSS velocity, consistent with both its morphology and the \citet[][]{2014karachentsevdisturber} association of this object with the NGC 4826 group.

In Figure \ref{fig:gasrichness-colour}, we show $\mhi/\LG$ for our \ion{H}{1} detections of UDGs (blue stars) and foreground dwarfs (green squares), and $\mhilim/\LG$ for our \ion{H}{1} non-detections (red downward arrows) as a function of $g-r$ (left panel) and $g-z$ (right panel).\ The vertical dashed lines in each panel show the median colors of the follow-up sample as a whole: $g-r=0.53$ and $g-z=0.79$.\ We note that our upper limits are generally higher than the $\mhi/\LG = 1 \msun/\lsun$ used to estimate the required integration times.\ There are three potential reasons for this reduction in sensitivity: integrations/scans flagged due to RFI, 20\% calibration adjustment, and/or noisier than expected data.\ By and large, our \ion{H}{1} detections have colors that are bluer than the \ion{H}{1} non-detections but the scatter is large (see also Figures \ref{fig:arcomp} - \ref{fig:dwcomp}); in the left panel, we also show the median $g-r=0.35$ of the entire HUDs sample as the vertical dashed-dotted line.\ Several of our \ion{H}{1} detections, including 6/9 UDGs, hover around this line and the vast majority of our non-detections lie on its redder side.\ We discuss the differences between the optical properties of our \ion{H}{1}-confirmed UDGs, foreground dwarfs, and \ion{H}{1} non-detections in Section \ref{subsec:gasrich}.

\section{Discussion} \label{sec:Discussion}
With our pilot sample of \ion{H}{1}-confirmed UDGs, foreground dwarfs, and \ion{H}{1} non-detections in hand, we provide some initial insight on three main questions our survey aims to answer: 1.\ Are there optical features that distinguish bona-fide gas-rich UDGs from foreground dwarfs or \ion{H}{1} non-detections among UDG candidates? 2.\ What constraints, if any, do our \ion{H}{1}-confirmed UDGs place on formation mechanisms? 3.\ How unusual are UDGs in the context of local galaxy scaling relations? We address these questions in Sections \ref{subsec:gasrich}, \ref{subsec:formation}, and \ref{subsec:btfr}, respectively.

\subsection{Comparing UDGs with \ion{H}{1} Detections and Non-detections} \label{subsec:gasrich}
Our follow-up \ion{H}{1} observations of 70 SMUDGes UDG candidates have revealed 9 gas-rich UDGs and 9 gas-rich foreground dwarf galaxies, while the remaining 52 targets were not detected in \ion{H}{1}.\ In this section, we explore differences between the environment and optical/NUV properties of these subsamples both to improve our detection efficiency in the full survey as well as to constrain the properties of \ion{H}{1}-rich and \ion{H}{1}-poor objects in the LSB regime.

We first revisit the spatial distribution of the follow-up targets shown in Figure \ref{fig:skydist}.\ The projected distribution of our sample spans both high and low density regions around Coma, with no obvious difference in location relative to the large-scale filamentary structure (grey circles) between \ion{H}{1} detections (blue stars and green squares) and non-detections (red circles).\ This qualitatively suggests that there is no strong correlation between \ion{H}{1} content and projected environment, implying that sky location is not a good predictor of gas richness among pilot sample galaxies.

Quantitatively, we find that none of the \ion{H}{1}-confirmed UDGs are likely to be gravitationally bound to the Coma Cluster based on their redshifts and projected spatial separations.\ Furthermore, only one of these objects (SMDG1248019+261236) has at least one massive companion $(M_g < -19\,\mathrm{mag})$ that projects within 300 kpc and within $\pm 500\,\kms$ (Kadowaki et al., in prep).\ While not obvious from Figure \ref{fig:skydist}, our \ion{H}{1}-confirmed UDGs reside in sparse environments.\ These findings are generally consistent with previous work that has investigated the environmental dependence of gas content \citep{2017browngasstripping}.

We next investigate whether or not discernible NUV emission in archival GALEX imaging predicts a detectable \ion{H}{1} reservoir among UDG candidates.\ The vast majority of pilot survey targets that are in the GALEX footprint do not have detectable NUV emission, which is commensurate with the findings of \citet{2019SMUDGesUV} for the broader SMUDGes sample.\ This is also the case for our \ion{H}{1} detections with GALEX All-sky Imaging Survey \cite[AIS;][]{2007MorrisseyGALEX,2009GALEX} coverage, raising the possibility that AIS-depth NUV imaging is not sufficient to detect ongoing star formation in UDGs.\ We therefore examine the subset of pilot survey targets with GALEX NUV exposures of at least 1000 seconds, i.e., Medium Imaging Survey (MIS) depth or $\sim$5-10 times deeper than the AIS.\ Of the 32 pilot survey targets in this category, 14 have discernible GALEX emission.\ All of these objects for which our \ion{H}{1} spectra are sensitive to at least $\mhi/\LG=2\,M_{\odot}/\lsun$ across the observed band have been detected in \ion{H}{1}, while many of the objects for which deep GALEX images reveal no NUV emission are \ion{H}{1} non-detections with $\mhilim/\LG< 1.5\,M_{\odot}/\lsun$.\ This suggests that MIS-depth NUV imaging is a good predictor of gas richness among SMUDGes UDG candidates.\

\begin{figure*}[!h]
\centering{}
\fboxsep=0.5mm
\fboxrule=0pt

\includegraphics[width=17.5cm]{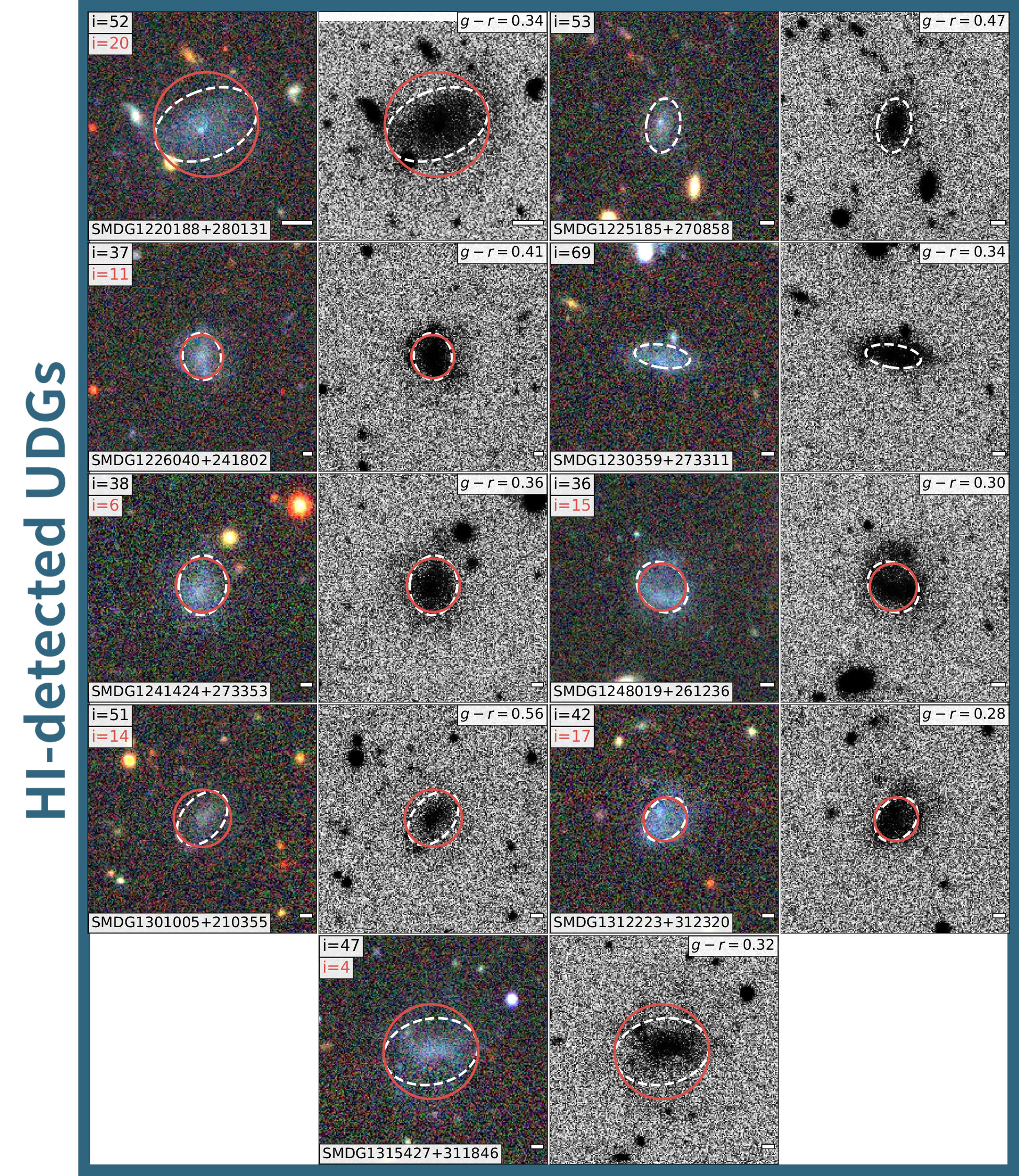}
\caption{$55\arcsec\times55\arcsec$ color and grayscale $grz$ image cutouts of \ion{H}{1}-detected UDGs shown in pairs with the color image on the left and the grayscale image on the right.\ The adjusted contrast and brightness of the color images highlights brighter emission in each object, while the histogram equalization of the grayscale images highlights the lower surface brightness emission.\ In all panels, the dashed, white ellipses have the disk geometry and semi-major axis, $\frac{a}{2}=r_{eff}$, of the best-fitting GALFIT models reported in Table \ref{table:maintable}.\ The object's color from Table \ref{table:maintable} is in the top-right corner of each image pair, and a scale bar that is 1 kpc across at the UDG distance is in the bottom-right corner.\ For a subset of the objects, we also overlay red ellipses corresponding to the disk geometry of GALFIT models with lower inclinations as detailed in Section \ref{subsec:btfr}.\ The inclinations of the corresponding disk, computed using Equation \ref{eqn:incl}, are in the top-left corner of each image pair.\ } 
\label{fig:arcomp}

\end{figure*}

\begin{figure*}[!h]
\centering{}
\includegraphics[width=17.5cm]{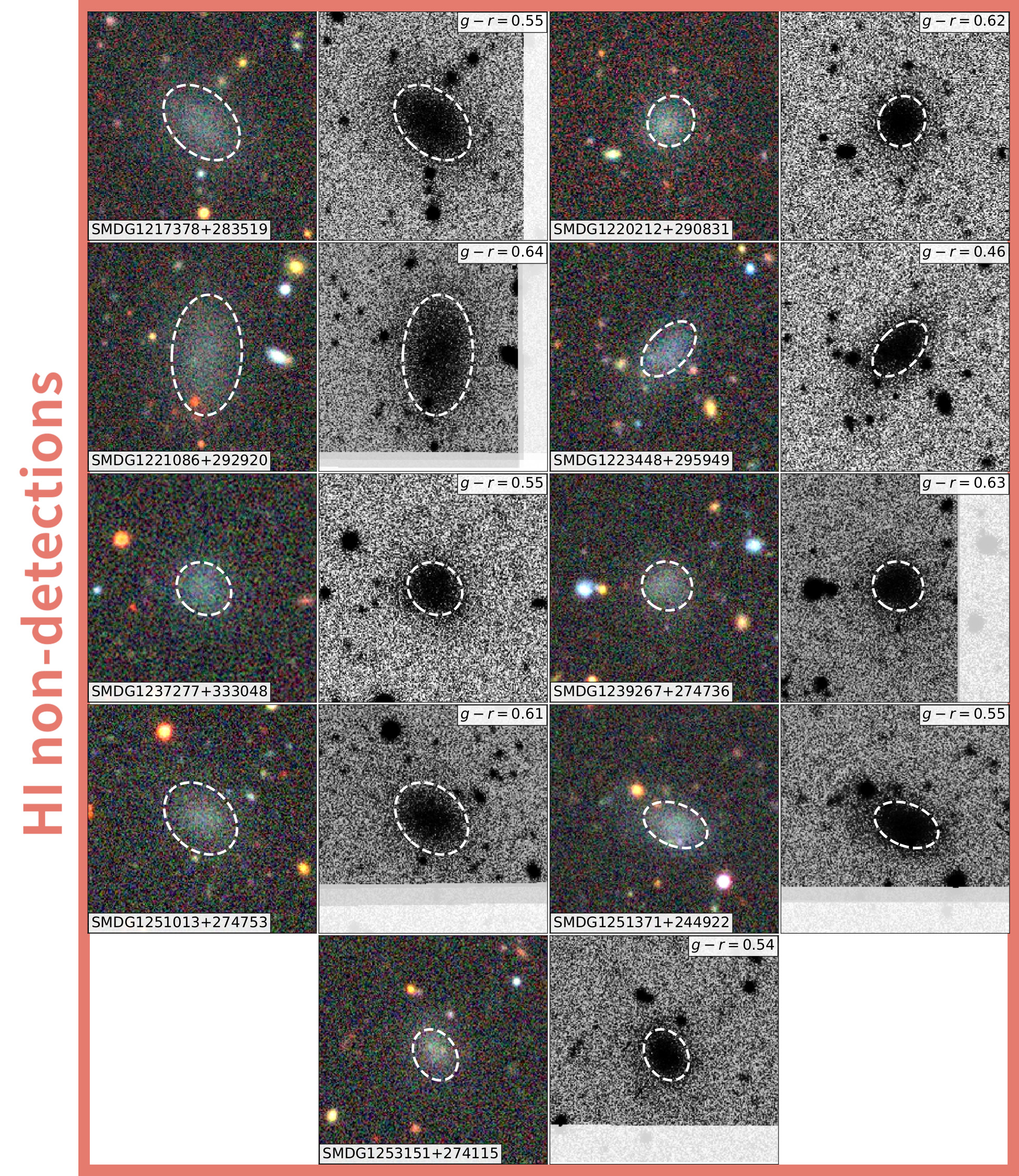}
\caption{Same as Figure \ref{fig:arcomp}, but for \ion{H}{1} non-detections.}
\label{fig:nondetectcomp}
\end{figure*}

\begin{figure*}[!h]
\centering{}
\includegraphics[width=17.5cm]{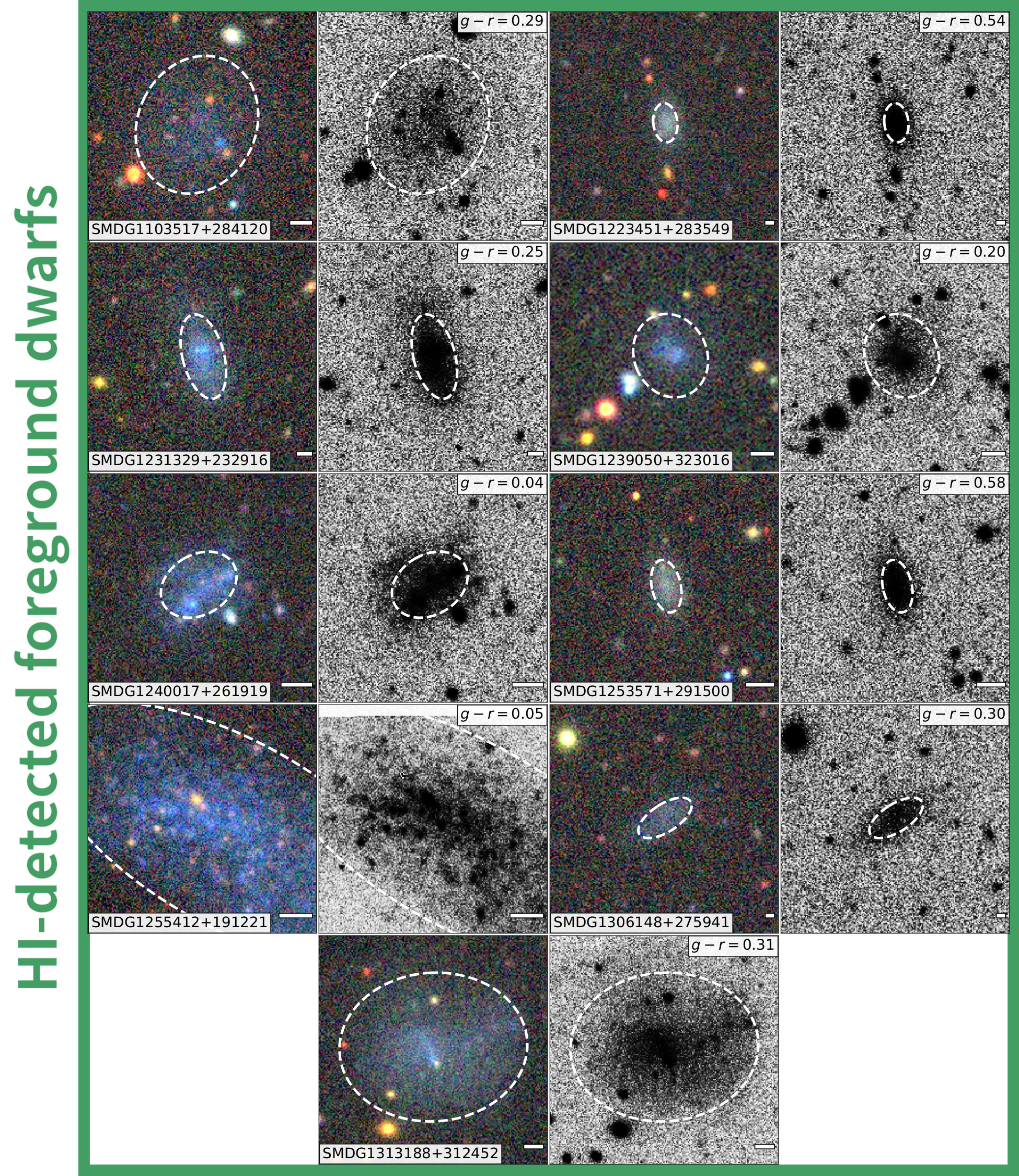}
\caption{Same as Figure \ref{fig:arcomp}, but for \ion{H}{1}-detected foreground dwarf galaxies and the scale bar represents 200 pc.}
\label{fig:dwcomp}
\end{figure*}

Finally, we compare the DECaLS optical morphologies of the \ion{H}{1}-confirmed UDGs, the UDG candidates that we did not detect in \ion{H}{1}, and the \ion{H}{1}-detected foreground dwarfs.\ Figures \ref{fig:arcomp}-\ref{fig:dwcomp} show color and grayscale $grz$ cutout pairs for these three subsets of our sample, where the adjusted contrast and brightness of the color image highlights the brighter emission in the object and the histogram equalization of the grayscale image highlights fainter emission.\ In all panels, the dashed, white ellipses have the disk geometry and semi-major axis, $\frac{a}{2}=r_{eff}$, of the best-fitting GALFIT models reported in Table \ref{table:maintable}.\ We note that Figures \ref{fig:arcomp} and \ref{fig:dwcomp} show all of the \ion{H}{1}-confirmed UDGs and foreground dwarfs in our sample, while Figure \ref{fig:nondetectcomp} shows images of a subset of the 52 \ion{H}{1} non-detections with similar $r_{eff}$ and $m_g$ to the \ion{H}{1}-confirmed UDGs.\ 

Figures \ref{fig:arcomp} and \ref{fig:nondetectcomp} demonstrate that on the whole, the \ion{H}{1}-detected UDGs are bluer than the UDG candidates that we do not detect, although as illustrated in Figure \ref{fig:gasrichness-colour} the scatter in color is large (c.f.\ SMDG1301005+210355 which we do detect in \ion{H}{1}, and SMDG1223448+295949 which we do not).\ This is consistent with the clear trends seen at higher surface brightnesses \citep[][]{huang2012galaxypop,GASS,Brown2015} as well as in other LSB studies \citep{UDGs-6Leisman,HSC-LSBs,2019prole}, suggesting that star formation proceeds similarly in high and low surface brightness galaxies \citep{2000Bell-LSB_SFH}.

Figures \ref{fig:arcomp} and \ref{fig:nondetectcomp} also illustrate that the \ion{H}{1}-detected UDGs are more irregular in morphology both within and beyond $r_{eff}$ than the undetected UDG candidates, although there is some scatter (c.f.\ SMDG1225185+270858 which we do detect in \ion{H}{1}, and SMDG1253151+274115 which we do not).\ On the other hand, the combination of DECaLS-depth color and morphology does appear to predict gas richness: blue and irregular objects in our pilot sample are almost invariably gas-rich, while red and smooth objects are invariably gas-poor.\  The efficiency of future \ion{H}{1} follow-up UDG campaigns can therefore be increased relative to the statistics presented here by preferentially targeting candidates that are both blue and irregular.

Do our \ion{H}{1}-confirmed UDGs differ in optical morphology from our gas-rich foreground dwarfs? Comparing Figures \ref{fig:arcomp} and \ref{fig:dwcomp} reveals that, among gas-rich objects, the foreground dwarfs tend to have larger angular sizes than the confirmed UDGs, consistent with Z19's hypothesis using a clustering analysis.\ The bluest gas-rich objects that we detect are also foreground dwarfs and not confirmed UDGs.\ While some stars in the very nearby dwarf SMDG1255412+191221 begin to appear resolved in the DECaLS imaging, we find no clear difference in optical morphology between bona-fide UDGs and foreground dwarfs in the pilot sample, making the two difficult to distinguish among follow-up targets.\ Blue foreground dwarfs are therefore an important potential contaminant among gas-rich UDG candidates identified by their optical colors and morphologies alone.\ Distance information is required to identify UDGs in the field.\

\subsection{Constraining Formation Mechanisms}\label{subsec:formation}
The stellar masses and velocity widths of our \ion{H}{1}-confirmed UDGs are commensurate with them being dwarf galaxies, in line with other estimates for UDGs in a variety of environments \citep[e.g.,][]{2018Sifon,PandyaUDG,SMUDGes,2020SPLUS}.\ How UDG-like field dwarfs could form is an active area of research (see Section \ref{sec:intro}), and the small size of the \ion{H}{1}-confirmed UDG sample from this pilot survey is too small for quantitative comparisons with theory.\ Nonetheless, we briefly consider the gas richnesses and sizes of the \ion{H}{1}-confirmed UDGs in the context of formation model predictions.

The star-formation feedback model presented by \citet{UDG-Formation2} predicts that UDGs in the field today have gas richnesses that scale with their sizes at fixed stellar mass.\ As shown in Figure \ref{fig:gasrichness-size}, we find evidence for a trend between $\mhi/M_*$ and $R_{eff}$ when the gas-rich UDGs are subdivided into two stellar mass bins.\ This trend persists when the gas-rich UDG samples from \citet[][who noted this trend in their smaller sample]{HCGUDGs} and \citet{2020ManceraPina} are also considered, but it is not evident in the foreground dwarf sample also plotted in Figure \ref{fig:gasrichness-size}.\ The correlation between gas richness and size for UDGs is qualitatively consistent with the predictions of \citet{UDG-Formation2}, although a similar trend may also emerge from other UDG formation scenarios.\ 

It is also possible that the correlations between gas richness and size exist in the broader galaxy population, and therefore that the trends in Figure \ref{fig:gasrichness-size} do not constrain UDG formation mechanisms at all.\ That the foreground dwarfs in our sample do not follow this trend argues against this possibility.\ Examining gas richnesses and sizes for a larger sample of galaxies might further clarify this issue, as might be obtained by homogenizing measured properties across the SPARC \citep{2016LelliSPARC}, SHIELD \citep{2011CannonSHIELD}, and LITTLE THINGS \citep{2012Hunter} samples along with samples of gas-rich UDGs.\ More data are needed to quantify comparisons between gas richnesses and sizes predicted by UDG formation models and other mechanisms, which we anticipate undertaking with data from the full survey.\

\subsection{Disk Geometry and the BTFR}\label{subsec:btfr}
We now discuss the \ion{H}{1}-confirmed UDGs in the context of the baryonic Tully-Fisher (BTFR) in order to explore the possibility that our sample exhibits an offset from this relation similar to that found by \citet{2019HUDsBTFR,2020ManceraPina}.\ Because our \ion{H}{1} detections stem from spatially unresolved single-dish observations, we must resort to optical measures of the disk geometry to estimate \ion{H}{1} disk rotation velocities, $V_{rot}$, from the measured velocity widths, $\wftyc$, in Tables \ref{table:detectiontable} and \ref{table:dwarftable}.\ We therefore proceed to derive $V_{rot}$ for our \ion{H}{1} detections, examine BTFR offsets in the context of the reliability with which we can estimate the disk geometry, and discuss the implications of these findings for UDG structure.

We first compute rotation velocities for our \ion{H}{1} detections using the relation for a flat axisymmetric disk:\begin{equation}\label{eqn:vrot}V_{rot}^{GF} = \frac{W_{50,c,t}}{2\mathrm{sin}(i^{GF})},\end{equation}where $W_{50,c,t}$ is the profile velocity width that has been corrected for ISM turbulence (see below) in addition to the instrumental effects discussed in Section \ref{subsec:detections} and $i^{GF}$ is the disk inclination implied by $b/a$ of the best-fitting GALFIT models of the optical UDG morphology given in Table \ref{table:maintable} and represented by the white ellipses in Figures \ref{fig:arcomp}.\  We calculate $i^{GF}$ via the standard relation: 
\begin{equation}\label{eqn:incl}\mathrm{cos^2}(i^{GF}) = \frac{(b/a)^2 - q^2_{0}}{1 - q^2_{0}} \, ,\end{equation}
where $q_0$ is the intrinsic axial ratio.\ We adopt $q_0 = 0.2$ in line with many previous studies \citep[e.g.][]{1994giovanelli,1997Giovanelli,UDGs-6Leisman}, although for our intermediate and low-inclination systems values as large as $q_0 =0.5$ \citep{2013Roychowdhury} only impact the derived $V_{rot}$ at the 10\% level.\ 

While the value of $q_{0}$ does not strongly impact the derived $V_{rot}$, we emphasize that there are considerable uncertainties in $i^{GF}$ derived from $b/a$ in Table \ref{table:maintable}.\ First, if the \ion{H}{1} disk is warped \citep[e.g.][]{2015Kamphuis} or if the \ion{H}{1} and optical disks are misaligned \citep[e.g.][]{2019Starkenburg, 2020ManceraPina}, $i^{GF}$ will not reflect the \ion{H}{1} disk geometry.\ Second, Figure \ref{fig:arcomp} illustrates that the \ion{H}{1}-confirmed UDGs have irregular morphologies, while the GALFIT models used to derive $b/a$ in Table \ref{table:maintable} assume a smooth distribution of light (Z19).\ This raises the possibility that clumps in the disk systematically pull $b/a$ away from the value that reflects the underlying disk geometry, biasing $i^{GF}$.\ We therefore consider $i^{GF}$ to be only a rough approximation of the \ion{H}{1} disk inclination that are much more uncertain than $b/a$ from the smooth GALFIT models listed in Table \ref{table:maintable}, and list them as such in Table \ref{table:incandrot}.\ We note that, since $d(sinx)/dx = cosx$ is much larger for low $x$ than when $x$ approaches $90^{\circ}$, uncertainties in $i^{GF}$ in low- and intermediate-inclination systems have a larger impact on $V_{rot}^{GF}$ than uncertainties on $i^{GF}$ in high-inclination systems.

We follow the prescription of \citet{2001turbulence} to correct $\wftyc$ in Tables \ref{table:detectiontable} and \ref{table:dwarftable} for ISM turbulence to obtain $W_{50,c,t}$, required in Equation \ref{eqn:vrot}, for the Gaussian profiles in Figure \ref{fig:detectspectra}:
\begin{equation}
    \label{eqn:turbcorr}
    \begin{split}
    W_{50,c,t} = W_{50,c}^2 + W_{T,50}^2[1-2\mathrm{e}^{-(\frac{W_{50,c}}{100})^2}] \\ 
    - 2W_{50,c}W_{T,50}[1-\mathrm{e}^{-(\frac{W_{50,c}}{100})^2}].\
    \end{split} 
\end{equation} 
The factor of $100\,\kms$ in the exponential terms accounts for the profile shapes at 50\% of their peak flux.\ We set $W_{T,50}= 5 \,\kms$ in Eq.\ \ref{eqn:turbcorr}, commensurate with estimates for systems with flat rotation curves by \citet{2001turbulence} and \citet{2012Kirby}, since dwarf galaxies rarely have declining rotation curves \citep{2006Cantinella,2016LelliSPARC}, and the UDG rotation curves from \citet{2020ManceraPina} are generally flat.\ We have also not attempted to correct for asymmetric drift in our unresolved data, although this may be significant for $V_{rot}^{GF} \lesssim 15 \,\kms$ \citep[e.g.][]{2017LITTLETHINGS,2017Read}.\ For these systems, $V_{rot}^{GF}$ is underestimated.\ If our \ion{H}{1} detections have rising rotation curves at the edges of their \ion{H}{1} disks as is the case for many dwarfs and some UDGs, then our choice of $W_{T,50}$ results in an over-correction.\ The resulting values of $V_{rot}^{GF}$ are given in Table \ref{table:incandrot}, which we consider highly uncertain due to the uncertainties in $i^{GF}$ discussed above.

\begin{figure*}[!h]
\includegraphics[width=18cm]{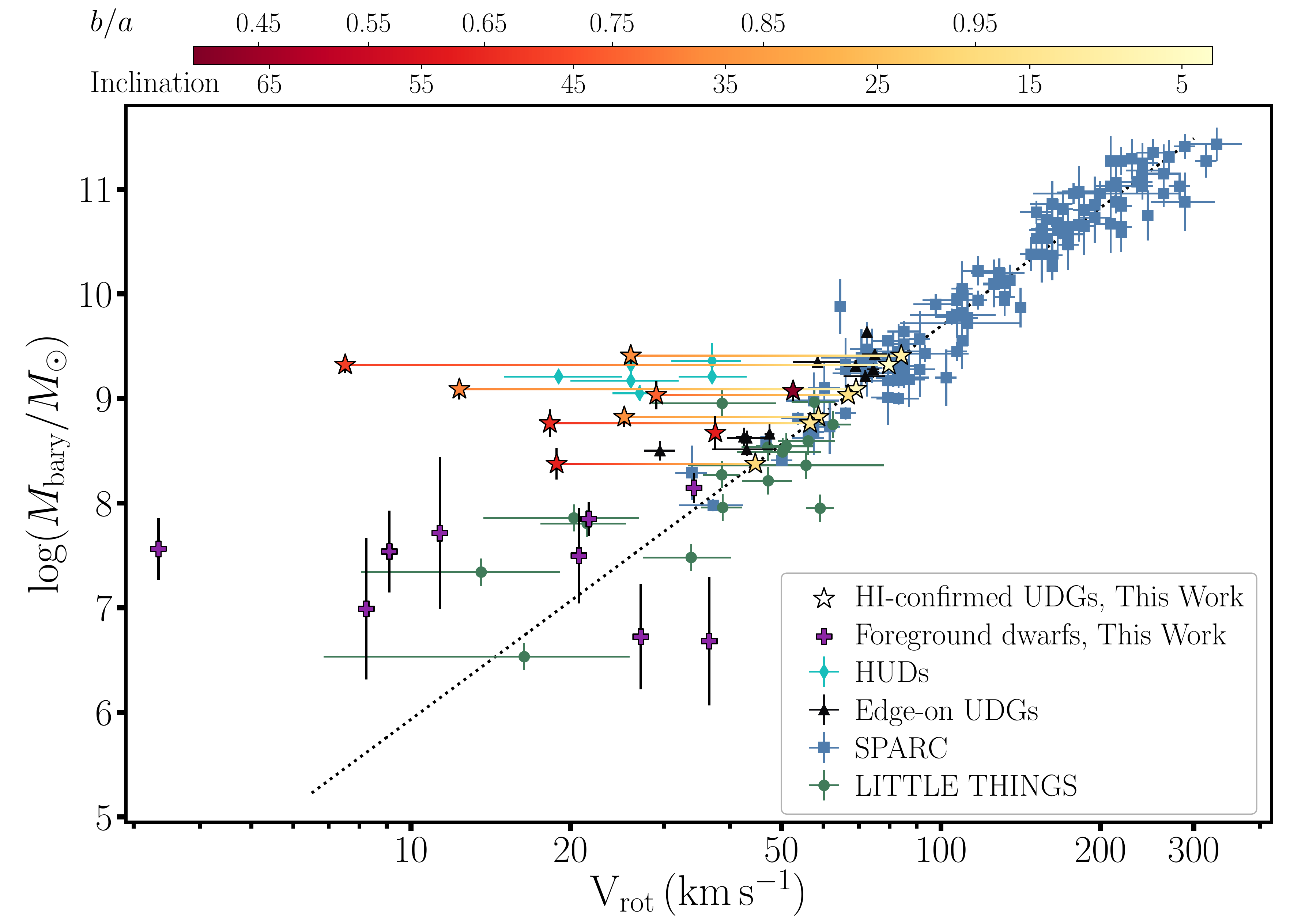}
\caption{Baryonic Tully-Fisher relation $(\mbary \mathrm{vs.}\, V_{rot})$ formed by the SPARC \citep[blue squares,][]{2016LelliSPARC} and LITTLE THINGS \citep[green circles,][]{2017LITTLETHINGS} samples, where $V_{rot}$ and $i$ are derived from standard tilted-ring kinematic modeling, with the best-fitting BTFR shown as a dotted black line.\ The 6 UDGs from the HUDs sample where $V_{rot}$ is estimated separately using a new method to determine $i$ \citep[cyan diamonds,][]{2019HUDsBTFR,2020ManceraPina} lie off the BTFR, while the edge-on HUDs \citep[black triangles,][]{2019He}, by and large, lie within its scatter.\ When we use the optically-derived inclinations, $i^{GF}$, and turbulence-corrected \ion{H}{1} linewidths, $W_{50,c,t}$, to estimate rotation velocities, $V_{rot}^{GF}$, 7/9 \ion{H}{1}-confirmed UDGs in our sample (orange and red stars) fall off this relation, as do some of our foreground dwarfs (purple crosses).\ The UDG symbols are colored according to their axial ratios/inclinations as shown in the colorbar.\ For the UDGs which fall off the BTFR, colored horizontal lines show how their axial ratios and inclinations change as they are brought onto the relation (from red to yellow).\ Representations of best-fit GALFIT models with the axial ratios corresponding to each pair of stars are shown overlaid on stacked optical images in Figure \ref{fig:arcomp}.\ It is plausible that the systematics of fitting smooth photometric models to clumpy, low inclination, LSB objects explains the offsets of the red stars from the BTFR.\ See text for details.\ }
\label{fig:btfr}
\end{figure*}

In Figure \ref{fig:btfr}, we show the BTFR composed of two samples of galaxies with spatially-resolved \ion{H}{1} maps: SPARC \citep[purple squares,][]{2016LelliSPARC} and LITTLE THINGS \citep[black circles,][]{2012Hunter,2017LITTLETHINGS}.\ In those samples, $V_{rot}$ has typically been measured using a standard tilted-ring approach \citep{1974Rogstad,1997Sicking} that fits for the disk geometry and rotation simultaneously to break the degeneracy between $V_{rot}$ and sin$i$ in the line-of-sight velocities.\ Figure \ref{fig:btfr} also shows the 6 intermediate-inclination UDGs from the HUDs sample which deviate from the BTFR\footnote{We have calculated the $M_{bary}$ and its uncertainties using the values from Table 1 of \citet{2020ManceraPina}.\ We note that the error bars in our Figure \ref{fig:btfr} for the UDGs from \citet{2019HUDsBTFR,2020ManceraPina} are smaller because they have propagated uncertainties in $M_{*}$ and $\mhi$ in logarithmic units instead of in linear units.} \citep{2019HUDsBTFR}, and the 11 edge-on (i.e.\ high-inclination) HUDs \citep{2019He} which by and large do not \citep{2020ManceraPina}.\ We note that, because the \ion{H}{1} maps kinematically modeled by \citet{2019HUDsBTFR,2020ManceraPina} do not have sufficient spatial resolution to constrain $V_{rot}$ and $i$ simultaneously \citep{2015diteodoro,2015Kamphuis}, a novel method where $i$ is estimated separately from $V_{rot}$ is adopted.\ On the other hand, any value of $i>75^{\circ}$ for the high-inclination UDGs of \citet{2019He} implies the same value of $V_{rot}$ since sin$i\sim1$.

The orange and red stars in Figure \ref{fig:btfr} show the locations of our \ion{H}{1}-confirmed UDGs in the $M_{bary}-V_{rot}$ plane when $V_{rot}^{GF}$ is used to estimate rotation velocities, with the symbol colour denoting $i^{GF}$ as given by the colorbar.\ The two UDGs with the largest $i^{GF}$ fall within the scatter of the relation defined by SPARC and LITTLE THINGS (dotted black line), while the rest do not.\ Given the uncertainties in $i^{GF}$ particularly at low inclinations, we calculate the inclinations $i^{BTFR}$ required to bring the discrepant points onto the BTFR, connecting pairs of stars corresponding to the same galaxy in Figure \ref{fig:btfr} with a horizontal line.\ These values of $i^{BTFR}$ are also given in Table \ref{table:incandrot}, the median $i^{BTFR}=14^{\circ}$.\ As expected from Equation \ref{eqn:vrot}, the discrepant points move on to the BTFR if the \ion{H}{1} disks of the corresponding UDGs have inclinations below $i^{GF}$.\ To constrain the plausibility with which a disk with $i^{BTFR}$ can reproduce the optical morphologies of the UDGs, we compute $(b/a)^{BTFR}$ implied by $i^{BTFR}$ using Equations \ref{eqn:vrot}-\ref{eqn:turbcorr} and overplot ellipses corresponding to the best-fitting GALFIT models obtained with $b/a=(b/a)^{BTFR}$ held fixed in red in Figure \ref{fig:arcomp}.

In light of the above considerations, we conclude that interpreting the available observations to mean that the \ion{H}{1}-confirmed SMUDGes UDGs deviate systematically from the BTFR is premature.\ The uncertainties in $i^{GF}$ are large, particularly in low-inclination systems.\ The white and red ellipses in Figure 6 demonstrate that in many cases, the GALFIT models that generated $i^{GF}$ and those produced holding $(b/a)^{BTFR}$ fixed produce nearly the same projected disk geometry.\ Furthermore, the irregular optical morphologies of the \ion{H}{1}-confirmed UDGs in Figure \ref{fig:arcomp} relative to our \ion{H}{1} non-detections evident in Figures \ref{fig:arcomp} and \ref{fig:nondetectcomp} raise the possibility that clumpy emission systematically biases the GALFIT fits that generated $i^{GF}$.\ Since there are few clumps in each object and since those clumps are rarely symmetrically distributed about the object center, it seems plausible that the effect of fitting these irregular LSB objects with smooth GALFIT models is to systematically under-estimate $b/a$ such that $i^{GF}$ is biased high.\ We emphasize that the SMUDGes UDG candidate selection criteria for low surface brightness and high ellipticity (Z19) favors low-inclination disks relative to high-inclination ones with the same $M_*$ and $R_eff$, and therefore that low-inclination disks should be over-represented in the SMUDGes sample compared to samples with a random distribution of sky orientations with a mean $i \sim 60^{\circ}$.\  Furthermore, since our \ion{H}{1} follow-up sample is effectively selected on luminosity (see Section \ref{sec:sample}) in a surface brightness-restricted sample, the objects in our sample are more likely still to be at low inclinations.\ It is therefore possible that most of the \ion{H}{1}-confirmed UDGs have low inclinations and that the $i^{GF}$ for those low-inclination systems is biased high.

A detailed investigation of potential biases in $i^{GF}$ for our \ion{H}{1}-confirmed UDGs is beyond the scope of this pilot paper, but we are carrying out simulations to quantify biases in smooth GALFIT models of irregular LSB galaxies as a function of their asymmetry \citep[e.g.,][]{1996Abraham,2003Abraham,2003Conselice} for the full survey.\ As a first check on our hypothesis, we estimate $V_{rot}^{GF}$ for the foreground dwarfs (which one would expect to lie within the scatter of the extrapolated BTFR, similar to other studies of the dwarf galaxy population; \citealt{2017LITTLETHINGS}; \citealt{2011CannonSHIELD}) and overplot them on Figure \ref{fig:btfr}.\ We emphasize that $V_{rot}^{GF}$ for both the foreground dwarfs and the UDGs are only order-of-magnitude estimates that assume the optical and \ion{H}{1} disks are aligned, and we do not attempt to quantify these significant uncertainties either in Table \ref{table:incandrot} or in Figure \ref{fig:btfr}.\ Nonetheless, at least some foreground dwarfs deviate from the BTFR similarly to the HI-confirmed UDGs, lending credence to our hypothesis that $i^{GF}$ is systematically overestimated.\ 

The gas-rich, intermediate-inclination UDG outliers from the BTFR studied by \citet{2019HUDsBTFR,2020ManceraPina} imply that the underlying structure and baryonic composition of these systems differs fundamentally from that assumed in any of the UDG formation scenarios posited so far (see Section \ref{sec:intro}).\ As proposed by these authors, a high stellar specific angular momentum, low star formation feedback scenario is one possible explanation.\ Examining Figure \ref{fig:btfr}, however, it is curious that the consistency of gas-rich UDG samples with the BTFR defined by higher surface brightness systems seems to depend on how their inclinations were measured: the edge-on systems studied by \citet{2019He} (where inclination uncertainties do not impact estimates of $V_{rot}$) are consistent with the BTFR, while the intermediate-inclination systems studied by \citet{2019HUDsBTFR,2020ManceraPina} (where $V_{rot}$ is measured independently from $i$ using a new technique) are outliers.\ The sensitivity of the locations of our low- and intermediate-inclination \ion{H}{1}-confirmed UDGs in the $M_{bary}-V_{rot}$ plane on the adopted inclination suggests that the effect of the viewing geometry should be carefully considered when inclination-dependent $V_{rot}$ are used to study the BTFR.

We emphasize that BTFR studies with SMUDGes UDGs that address the possible inclination dependence of offsets from this relation require \ion{H}{1} imaging with sufficient angular and spectral resolution to simultaneously model $V_{rot}$ and $i$ using standard tilted ring approaches.\ This is feasible for a small subset of the \ion{H}{1} detections presented here, and work in this regard is underway.

\section{Conclusions} \label{sec:Conclusion}
We have presented GBT \ion{H}{1} observations of 70 optically-detected SMUDGes UDG candidates (Z19) with $m_g \lesssim 19.5\,\mathrm{mag}$ in the Coma region.\ We detect \ion{H}{1} reservoirs in 18 of them (Figure \ref{fig:detectspectra}), measuring systemic velocities, $\vsys$, velocity widths, $\wftyc$, and flux integrals, $\int S dv$, directly from the spectra.\ Using kinematic distances estimated from $\vsys$, we compute \ion{H}{1} masses, $\mhi$, from the spectra as well as stellar masses, $M_{*}$, and half-light radii, $R_{eff}$, from GALFIT models to the deep DECaLS imaging.\ We use $R_{eff}$ to confirm that 9 of our \ion{H}{1} detections satisfy the size criterion defining UDGs, while the remainder are foreground dwarfs (Tables \ref{table:detectiontable} and \ref{table:dwarftable}).\ Although only a pilot for a much larger GBT program that is currently underway, these observations already represent the largest \ion{H}{1} follow-up campaign of optically-selected UDG candidates ever reported, and the 9 confirmed UDGs are the largest available sample of optically-selected UDGs with \ion{H}{1} detections.

Comparing the properties of our \ion{H}{1}-detected UDGs, \ion{H}{1}-detected foreground dwarfs and our \ion{H}{1} non-detections, we find similar sky distributions relative to the Coma large-scale structure (Figure \ref{fig:skydist}) but that 8/9 UDGs are in low-density environments with no massive ($M_g< -19\,\mathrm{mag}$) companions within $R_{proj}= 300\,$kpc or $\Delta \vsys = \pm 500\,\kms$.\ In addition, our \ion{H}{1} detections typically have counterparts in the NUV if the exposures are sufficiently deep ($\gtrsim\,1000\,$sec with GALEX).\ In DECaLS-depth optical imaging, the gas-rich UDGs are bluer and smoother in morphology than the UDG candidates that we do not detect in \ion{H}{1} but the scatter is large in both properties (Figures \ref{fig:gasrichness-colour}, \ref{fig:arcomp}, and \ref{fig:nondetectcomp}).\ On the other hand, targets that are both blue and irregular are gas-rich, while those that are both red and smooth are gas-poor: it is the combination of optical morphology and color that best predicts gas richness.\ Although the angular sizes of the foreground dwarfs are typically larger than those of the \ion{H}{1}-confirmed UDGs, there is little difference in optical morphology or color between these subsamples (Figures \ref{fig:arcomp} and \ref{fig:dwcomp}).\ Without distance information, foreground dwarfs contaminate samples of optically blue, irregular UDG candidates.

Commensurate with tentative results for blue UDGs around galaxy groups \citep{HCGUDGs}, we find evidence for a correlation between the gas richness, $\mhi/M_{*}$, and size, $R_{eff}$,  when our \ion{H}{1}-confirmed UDGs as well as other gas-rich UDGs are divided into two stellar mass bins (Figure \ref{fig:gasrichness-size}).\ The same trend is not obvious for the foreground dwarfs.\ The correlation between UDG gas richness and size suggested by the data is broadly consistent with predictions from the star formation feedback model for UDG formation \citep{UDG-Formation2}, although other mechanisms may also produce the trend.

We place our \ion{H}{1}-confirmed UDGs on the BTFR using best-fitting inclinations, $i^{GF}$, from smooth GALFIT models of DECaLS imaging and turbulence-corrected velocity widths to estimate rotation velocities $V_{rot}^{GF}$.\ We find that the 7/9 objects with the lowest $i^{GF}$ have lower $V_{rot}^{GF}$ than expected from the BTFR defined by high surface brightness, gas-rich galaxies with \ion{H}{1} rotation curves and disk geometries derived from kinematic models (Figure \ref{fig:btfr}), similar to that found by \citet{2019HUDsBTFR,2020ManceraPina} for a sample of marginally-resolved HUDs using a new technique for constraining $i$ separately from $V_{rot}$ via the \ion{H}{1} morphology.\ For our sample, however, we find that plausible systematics resulting from the application of smooth GALFIT models to clumpy, low-inclination LSB objects are sufficient to reconcile these discrepancies (Figures \ref{fig:arcomp} and \ref{fig:btfr}) precluding a meaningful analysis of BTFR offsets.\ We plan on investigating this trend and its implications in detail with our full follow-up sample.\ 

The pilot survey results presented here provide some initial insight into the properties of gas-rich UDGs and the mechanisms by which they form.\ Despite being the largest of its kind, our sample of confirmed gas-rich optically-detected UDGs remains small.\ A much larger SMUDGes \ion{H}{1} follow-up campaign is underway at the GBT.\ We ultimately plan on targeting over 200 objects, and expect to confirm at least 50 gas-rich UDGs.\ This larger sample will enable quantitative investigations of the interplay between gas richness and UDG properties in order to understand how they form and evolve.\ Furthermore, it will also provide predictive insight into the gas properties of UDG candidates in the eventual $\sim10,000\,\mathrm{deg^2}$ SMUDGes survey.\

\floattable
\begin{deluxetable}{cccCCCCCCCCCCcc}[!h]
\tablecaption{Target UDG Candidate Properties \label{table:maintable}}
\rotate
\tabletypesize{\tiny}
\tablehead{
\colhead{Name} & \colhead{RA} & \colhead{Dec} & \colhead{$m_g$} & \colhead{$\mu_{0,g}$} & \colhead{$g-r$} & \colhead{$g-z$} & \colhead{$r_{eff}$} & \colhead{$b/a$} & \colhead{${\theta}$} & \colhead{$n$} & \colhead{Int.\ Time} & \colhead{${\sigma_{50}}$} & \colhead{Ref} & \colhead{HI} \\
\colhead{} & \colhead{H:M:S} & \colhead{D:M:S} &  \colhead{(mag)} & \colhead{${(\mathrm{\frac{mag}{arcsec^{2}}})}$} & \colhead{(mag)} & \colhead{(mag)} & \colhead{(arcsec)} &\colhead{} & \colhead{(deg)} & \colhead{} & \colhead{(hours)} & \colhead{(mJy)} & \colhead{} & \colhead{Det?}\\
\colhead{(1)} & \colhead{(2)} & \colhead{(3)} & \colhead{(4)} & \colhead{(5)} & \colhead{(6)} & \colhead{(7)} & \colhead{(8)} & \colhead{(9)} & \colhead{(10)} & \colhead{(11)} & \colhead{(12)} & \colhead{(13)} & \colhead{(14)} & \colhead{(15)}
} 
\startdata
SMDG1103517+284120 & 11:03:51.7 & 28:41:20 & 18.06\pm0.01 & 25.25\pm0.14 & 0.32\pm0.02 & 0.20\pm0.07 & 16.2 \pm 0.9 & 0.83 \pm 0.01 & -29 \pm 3 & 0.71 \pm 0.04 & 0.2 & 1.39 & K20 (M98) & Y\\
SMDG1217378+283519 & 12:17:38.0 & 28:35:20 & 18.98\pm0.01 & 24.95\pm0.05 & 0.54\pm0.01 & 0.87\pm0.02 & 10.2 \pm 0.2 & 0.70 \pm 0.01 & 47 \pm 1 & 0.73 \pm 0.02 & 0.5 & 0.58 & Z19 & \\
SMDG1217443+332043 & 12:17:44.2 & 33:20:44 & 18.49\pm0.01 & 25.99\pm0.10 & 0.46\pm0.03 & 0.60\pm0.06 & 17.1 \pm 0.7 & 0.77 \pm 0.01 & 84 \pm 2 & 0.52 \pm 0.03 & 0.5 & 0.69 & Z19 & \\
SMDG1217451+281724 & 12:17:45.0 & 28:17:25 & 20.04\pm0.04 & 25.71\pm0.12 & 0.66\pm0.05 & 1.11\pm0.08 & 7.3 \pm 0.3 & 0.82 \pm 0.03 & 0 \pm 7 & 0.56 \pm 0.04 & 2.4 & 0.38 & Z19 & \\
SMDG1220188+280131 & 12:20:19.0 & 28:01:34 & 18.44\pm0.01 & 24.38\pm0.09 & 0.34\pm0.01 & 0.50\pm0.03 & 12.7 \pm 0.4 & 0.63 \pm 0.01 & -67 \pm 1 & 0.96 \pm 0.03 & 0.2 & 0.79 & K20 (H18) & Y\\
SMDG1220212+290831 & 12:20:21.0 & 29:08:34 & 19.48\pm0.01 & 24.30\pm0.06 & 0.62\pm0.02 & 0.88\pm0.04 & 6.1 \pm 0.1 & 0.90 \pm 0.01 & -24 \pm 5 & 0.92 \pm 0.02 & 1.3 & 0.41 & Z19\\
SMDG1221086+292920 & 12:21:17.0* & 29:29:21 & 18.77\pm0.01 & 25.27\pm0.06 & 0.64\pm0.01 & 1.03\pm0.02 & 14.1 \pm 0.3 & 0.57 \pm 0.01 & -4 \pm 1 & 0.70 \pm 0.02 & 0.4 & 0.73 & Z19\\
SMDG1221235+303643 & 12:21:23.4 & 30:36:44 & 19.01\pm0.01 & 25.55\pm0.11 & 0.54\pm0.02 & 0.68\pm0.04 & 14.4 \pm 0.6 & 0.63 \pm 0.01 & 65 \pm 1 & 0.76 \pm 0.04 & 0.6 & 0.84 & Z19\\
SMDG1221401+284346 & 12:21:40.0 & 28:43:47 & 19.26\pm0.01 & 25.00\pm0.06 & 0.64\pm0.02 & 0.91\pm0.03 & 9.5 \pm 0.2 & 0.60 \pm 0.01 & 77 \pm 1 & 0.67 \pm 0.02 & 0.9 & 0.53 & Z19\\
SMDG1221497+283111 & 12:21:49.7 & 28:31:12 & 19.00\pm0.03 & 25.83\pm0.13 & 0.59\pm0.04 & 0.82\pm0.05 & 15.0 \pm 0.7 & 0.64 \pm 0.02 & 9 \pm 2 & 0.65 \pm 0.04 & 0.5 & 0.85 & Z19\\
\enddata

\tablecomments{The first 10 rows of this table are shown here.\ The full table is available online in machine readable format.\ \\ \\ col.(1): Adopted SMUDGes UDG candidate name.\ cols.(2) and (3): J2000 position of optical centroid, which corresponds to our GBT LOS.\ RA values with an asterisk (*) indicate an offset (in RA and/or Dec) in the GBT pointing position.\ cols.(4) and (5): $g-$band apparent magnitude and central surface brightness.\ cols.(6) and (7): $g-r$ and $g-z$ colors.\ col.\ (8)-(11): Best-fitting effective radius, axial ratio, position angle, and $\mathrm{S\acute{e}rsic}$ index in GALFIT model of UDG candidate.\  col.(12): Total effective GBT integration time, including the ON+OFF positions and subtracting any time lost due to RFI.\ col.(13): Representative RMS noise of the spectrum at a velocity resolution of $\Delta V = 50 \, \kms$.\  col.(14): Reference from which UDG candidate is selected, alternative references in parentheses.\ S97 = \citet{1997Schombert}; M98 = \citet{1998Martin}; H18 = \citet{Haynes2018}.\ col.(15): \ion{H}{1} detection?}
\end{deluxetable}

\setlength{\tabcolsep}{3pt}
\begin{deluxetable*}{cCCCCCCCCCC}[htb!]
\caption{Properties of UDG with \ion{H}{1} detections}
\label{table:detectiontable}
\tablehead{
\colhead{Name} & \colhead{$\Delta V$} & \colhead{$\sigma_{\Delta V}$} & \colhead{$V_{sys}$} & \colhead{$\wftyc$} & \colhead{$S_{HI}$} & \colhead{$D_{HI}$} & \colhead{log($\mhi$)} & \colhead{log($M_{*}$)} & \colhead{log$\mbary$} & \colhead{$R_{eff}$} \\
\colhead{} & \colhead{($\kms$)} & \colhead{(mJy)} & \colhead{($\kms$)} & \colhead{($\kms$)} & \colhead{(Jy$\,\kms$)} & \colhead{(Mpc)} & \colhead{(log[$\msun$])} & \colhead{(log[$\msun$])} & \colhead{(log[$\msun$])} & \colhead{(kpc)} \\
\colhead{(1)} & \colhead{(2)} & \colhead{(3)} & \colhead{(4)} & \colhead{(5)} & \colhead{(6)} & \colhead{(7)} & \colhead{(8)} & \colhead{(9)} & \colhead{(10)} & \colhead{(11)}
}
\startdata
SMDG1220188+280131 & 10 & 1.86 & 2283\pm2 & 32\pm2 & 0.61\pm0.10 & 32.6 & 8.19\pm0.15 & 7.51\pm0.24 & 8.37\pm0.15 & 2.00\pm0.30 \\
SMDG1225185+270858 & 20 & 0.69 & 5888\pm5 & 63\pm7 & 0.17\pm0.07 & 84.1 & 8.44\pm0.19 & 8.01\pm0.21 & 8.67\pm0.16 & 2.63\pm0.18  \\
SMDG1226040+241802 & 15 & 0.66 & 11585\pm3 & 34\pm4 & 0.26\pm0.05 & 165.5 & 9.22\pm0.08 & 8.57\pm0.20 & 9.41\pm0.08 & 4.48\pm0.17 \\
SMDG1230359+273311 & 25 & 0.56 & 6794\pm3 & 102\pm5 & 0.37\pm0.08 & 97.1 & 8.91\pm0.10 & 8.01\pm0.20 & 9.07\pm0.10 & 3.11\pm0.17 \\
SMDG1241424+273353 & 10 & 1.52 & 7766\pm2 & 17\pm2 & 0.26\pm0.06 & 110.9 & 8.87\pm0.11 & 8.36\pm0.20 & 9.09\pm0.10 & 3.81\pm0.20 \\
SMDG1248019+261236 & 25 & 0.51 & 6043\pm5 & 32\pm7 & 0.24\pm0.05 & 86.3 & 8.63\pm0.10 & 8.01\pm0.21 & 8.82\pm0.09 & 2.76\pm0.17 \\
SMDG1301005+210355 & 25 & 0.53 & 7051\pm3 & 31\pm5 & 0.13\pm0.05 & 100.7 & 8.49\pm0.16 & 8.22\pm0.20 & 8.76\pm0.13 & 3.57\pm0.28 \\
SMDG1312223+312320 & 20 & 1.05 & 7487\pm3 & 42\pm4 & 0.26\pm0.09 & 107.0 & 8.84\pm0.15 & 8.18\pm0.20 & 9.03\pm0.13 & 3.00\pm0.15 \\
SMDG1315427+311846 & 25 & 0.99 & 7486\pm6 & 13\pm8 & 0.51\pm0.08 & 106.9 & 9.14\pm0.08 & 8.44\pm0.20 & 9.32\pm0.08 & 5.85\pm0.41 \\
\enddata
\tablecomments{col.(2): Velocity resolution of spectrum used to compute \ion{H}{1} properties (see Figure \ref{fig:detectspectra}).\ col.(3): RMS noise of spectrum at $\Delta V$ in col.(2).\ col.(4): Heliocentric systemic velocity.\ col.(5): Velocity width of the \ion{H}{1} detection, corrected for cosmological redshift and instrumental broadening.\ col.(6): Integrated \ion{H}{1} flux.\ col.(7): Distance estimated using the Hubble-Lema\^{i}tre Law, $\vsys$ and $\mathrm{H}_0 = 70 \, \kms\,\mathrm{Mpc}^{-1}$.\ We adopt distance uncertainties of 5 Mpc.\ col.(8): Logarithm of \ion{H}{1} mass calculated from Eq.\ref{eqn:himass} using $S_{HI}$ in col.(6) and $D_{HI}$ in col.(7).\ col.(9): Logarithm of stellar mass calculated using $m_g$ and $g-r$ from Table \ref{table:maintable}, $D_{HI}$ in col.(7), and the corresponding relation from \citet{2017zhang}.\ col.(10): Logarithm of baryonic mass, $1.33\mhi + M_{*}$.\ col.(11): Effective radius in physical units using $r_{eff}$ from Table \ref{table:maintable} and $D_{HI}$ in col.(7).\ }
\end{deluxetable*}

\begin{deluxetable*}{cCCCCCCCCCC}[htb!]
\caption{\ion{H}{1} Properties of Dwarfs}
\label{table:dwarftable}
\tablehead{
\colhead{Name} & \colhead{$\Delta V$} & \colhead{$\sigma_{\Delta V}$} & \colhead{$V_{sys}$} & \colhead{$W_{50,c}$} & \colhead{$S_{HI}$} & \colhead{$D_{HI}$} & \colhead{log($\mhi$)} & \colhead{log($M_{*}$)} & \colhead{log$\mbary$} & \colhead{$R_{eff}$}\\
\colhead{} & \colhead{($\kms$)} & \colhead{(mJy)} & \colhead{($\kms$)} & \colhead{($\kms$)} & \colhead{(Jy$\,\kms$)} & \colhead{(Mpc)} & \colhead{(log[$\msun$])} & \colhead{(log[$\msun$])} & \colhead{(log[$\msun$])} & \colhead{(kpc)} \\
\colhead{(1)} & \colhead{(2)} & \colhead{(3)} & \colhead{(4)} & \colhead{(5)} & \colhead{(6)} & \colhead{(7)} & \colhead{(8)} & \colhead{(9)} & \colhead{(10)} & \colhead{(11)}
}
\startdata
SMDG1103517+284120 & 10 & 2.46 & 668\pm2 & 26\pm3 & 0.97\pm0.13 & 9.5 & 7.32\pm0.45 & 6.58\pm0.50 & 7.49\pm0.46 & 0.75\pm0.40\\
SMDG1223451+283549 & 20 & 0.51 & 2377\pm4 & 59\pm5 & 0.34\pm0.05 & 34.0 & 7.97\pm0.15 & 7.19\pm0.23 & 8.14\pm0.14 & 0.78\pm0.02\\
SMDG1231329+232916 & 25 & 0.82 & 1060\pm6 & 8\pm8 & 0.44\pm0.07 & 15.1 & 7.38\pm0.29 & 6.68\pm0.35 & 7.56\pm0.30 & 0.77\pm0.02\\
SMDG1239050+323016 & 10 & 0.82 & 613\pm1 & 32\pm2 & 0.19\pm0.04 & 8.8 & 6.54\pm0.50 & 5.79\pm0.53 & 6.72\pm0.50 & 0.42\pm0.24\\
SMDG1240017+261919 & 10 & 2.24 & 452\pm2 & 13\pm3 & 0.72\pm0.10 & 6.5 & 6.85\pm0.68 & 5.64\pm0.70 & 6.99\pm0.67 & 0.30\pm0.23\\
SMDG1253571+291500 & 25 & 0.45 & 502\pm4 & 67\pm5 & 0.23\pm0.05 & 7.2 & 6.45\pm0.61 & 6.02\pm0.64 & 6.68\pm0.61 & 0.22\pm0.15\\
SMDG1255412+191221 & 5 & 4.76 & 420\pm1 & 19\pm2 & 4.37\pm0.21 & 6.0 & 7.57\pm0.72 & 6.38\pm0.75 & 7.71\pm0.72 & 0.75\pm0.62\\
SMDG1306148+275941 & 15 & 0.55 & 2559\pm3 & 42\pm4 & 0.15\pm0.04 & 36.6 & 7.67\pm0.17 & 6.94\pm0.23 & 7.84\pm0.16 & 1.28\pm0.17\\
SMDG1313188+312452 & 25 & 1.25 & 802\pm5 & 13\pm8 & 0.50\pm0.10 & 11.5 & 7.19\pm0.39 & 7.14\pm0.42 & 7.53\pm0.39 & 1.24\pm0.54\\\enddata
\centering
\tablecomments{All parameters have the same definitions as in Table \ref{table:detectiontable}.\ }
\end{deluxetable*}

\begin{deluxetable}{cCCC}
\caption{ \ion{H}{1} Properties of Non-detections}
\label{table:upperlimits}
\tablehead{
\colhead{Name} & \colhead{$D_{lim}$} & \colhead{log$(\mhilim)$} & \colhead{$({\mhilim}/{\LG})$}  \\
\colhead{} & \colhead{(Mpc)}& \colhead{(log[$\msun$])} & \colhead{$({\msun}/{\lsun})$} \\
\colhead{(1)} & \colhead{(2)} & \colhead{(3)} & \colhead{(4)}
}
\startdata
SMDG1217378+283519 & 7^{a} & 6.22 &  1.27\\
SMDG1217443+332043 & 100 & 8.61 &  0.96\\
SMDG1217451+281724 & 100 & 8.35 &  2.22\\
SMDG1220212+290831 & 100 & 8.39 &  1.43\\
SMDG1221086+292920* & 14.6^{a} & 6.96 &  1.32\\
SMDG1221235+303643 & 100 & 8.69 &  1.89\\
SMDG1221401+284346 & 100 & 8.49 &  1.50\\
SMDG1221497+283111 & 100 & 8.7 &  1.92\\
SMDG1221577+281436 & 100 & 8.8 &  1.06\\
SMDG1223448+295949 & 100 & 8.54 &  1.58\\
SMDG1224082+280544 & 100 & 8.41 &  0.82\\
SMDG1224166+291506 & 100 & 8.18 &  1.49\\
SMDG1225265+311646* & 100 & 8.59 &  1.19\\
SMDG1231418+264433 & 100 & 8.3 &  2.49\\
SMDG1237277+333048 & 100 & 8.33 &  1.19\\
SMDG1239267+274736 & 100 & 8.41 &  1.81\\
SMDG1239503+244949 & 100 & 8.53 &  1.36\\
SMDG1240119+251447 & 100 & 8.4 &  2.76\\
SMDG1247233+180140 & 100 & 8.69 &  1.02\\
SMDG1249413+270645 & 100 & 8.27 &  1.49\\
SMDG1251013+274753* & 87^{a} & 8.5 &  1.86\\
SMDG1253048+253121 & 100 & 8.44 &  1.97\\
SMDG1253151+274115* & 100^{b} & 8.28 &  1.63\\
SMDG1253489+273934 & 100 & 8.79 &  2.86\\
SMDG1254252+194332 & 100 & 8.61 &  0.83\\
SMDG1254556+285846 & 100 & 8.45 &  1.89\\
SMDG1255336+213035 & 100 & 8.12 &  1.59\\
SMDG1307464+291230 & 100 & 8.44 &  2.51\\
SMDG1308296+271354 & 100 & 8.35 &  1.98\\
SMDG1322561+314804 & 100 & 8.61 &  2.06\\
SMDG1226306+220532 & 100 & 8.63 &  0.95\\
SMDG1231070+253508* & 100 & 8.62 &  2.28\\
SMDG1232244+274043 & 100 & 8.22 &  1.46\\
SMDG1233516+234545 & 100 & 8.49 &  0.73\\
SMDG1234503+293313 & 100 & 8.49 &  1.34\\
SMDG1235065+263342 & 100 & 8.27 &  1.57\\
SMDG1240490+254406 & 100 & 8.45 &  3.50\\
SMDG1241097+221223 & 100 & 8.33 &  1.92\\
SMDG1245022+230956 & 100 & 8.28 &  1.44\\
SMDG1246029+255724 & 100 & 8.51 &  1.54\\
SMDG1248202+183824 & 100 & 8.7 &  0.37\\
SMDG1249353+253106 & 100 & 8.56 &  0.72\\
SMDG1251291+284433* & 100 & 8.29 &  2.80\\
SMDG1251337+314240 & 100 & 8.44 &  1.12\\
SMDG1251371+244922 & 100 & 8.46 &  1.28\\
SMDG1252056+221556 & 100 & 8.2 &  1.71\\
SMDG1252402+262602* & 100 & 8.48 &  0.85\\
SMDG1302417+215954 & 3.8^{c} & 5.94 &  0.25\\
SMDG1306158+273459 & 100 & 8.35 &  2.32\\
SMDG1312226+195525 & 100 & 8.27 &  1.68\\
SMDG1322538+220445* & 100 & 8.4 &  1.72\\
SMDG1333509+275006 & 100 & 8.38 &  1.72\\
\enddata
$^{\mathrm{a}}${Kadowaki et al., in prep},$^{\mathrm{b}}${\citet{Kadowaki2017}},$^{\mathrm{c}}${\citet{2014KimExVirgo}} \tablecomments{col.\ (2): Adopted distance in Eq.\ref{eqn:hilim}; see text for details.\ col.\ (3): $5\sigma$ upper limit on $\mhi$ calculated from Eq.\ref{eqn:hilim} using $D_{lim}$ from col.\ (2) and $\sigma_{50}$ from Table \ref{table:maintable}.\ col.(4): Upper limit on the gas richness (which is distance independent).}
\end{deluxetable}

\begin{deluxetable}{cCCC}
\caption{ Inclinations and Rotation Velocities}
\label{table:incandrot}
\tablehead{
\colhead{Name} & \colhead{$i^{GF}$} & \colhead{$V_{rot}^{GF}$} & \colhead{$i^{BTFR}$} \\
\colhead{} & \colhead{(deg)} & \colhead{(${\kms}$)} & \colhead{(deg)} \\
\colhead{(1)} & \colhead{(2)} & \colhead{(3)} & \colhead{(4)}
}
\startdata
\textbf{\ion{H}{1}-confirmed UDGs} \\
SMDG1220188+280131  & \sim 52 & \sim 19 & \sim 20 \\
SMDG1225185+270858  & \sim 53 & \sim 38 &  -  \\
SMDG1226040+241802  & \sim 37 & \sim 26 & \sim 11 \\
SMDG1230359+273311  & \sim 69 & \sim 53 &  - \\
SMDG1241424+273353  & \sim 38 & \sim 12 & \sim 6 \\
SMDG1248019+261236  & \sim 36 & \sim 25 & \sim 15 \\
SMDG1301005+210355  & \sim 51 & \sim 18 & \sim 14 \\
SMDG1312223+312320  & \sim 42 & \sim 29 & \sim 17 \\
SMDG1315427+311846  & \sim 47 & \sim 8 & \sim 4 \\
\hline 
\textbf{Foreground dwarfs} \\
SMDG1103517+284120  & \sim 35 & \sim 21 &  - \\
SMDG1223451+283549  & \sim 54 & \sim 34 &  - \\
SMDG1231329+232916  & \sim 63 & \sim 3 &  \sim 6  \\
SMDG1239050+323016  & \sim 33 & \sim 27 &  - \\
SMDG1240017+261919  & \sim 43 & \sim 8 &  \sim 17  \\
SMDG1253571+291500  & \sim 60 & \sim 37 & - \\
SMDG1255412+191221  & \sim 49 & \sim 11 &  \sim 17 \\
SMDG1306148+275941  & \sim 65 & \sim 22 &  - \\
SMDG1313188+312452  & \sim 39 & \sim 9 &  \sim 12  \\
\enddata
\tablecomments{col.(2): Inclination calculated using Eq.\ref{eqn:incl}, $b/a$ from Table \ref{table:maintable}, and an intrinsic axial ratio of $q_0=0.2$.\ col.(3): Rotational velocity calculated using $\wftyc$ corrected for turbulence and $i$ in col.(2).\ Given the systematics associated with measuring inclinations of clumpy low-inclination objects from smooth models, we consider $i^{GF}$ and $V_{rot}^{GF}$ to be rough estimates (see text).\ cols.(4): Inclinations required to lie on the BTFR for UDGs and dwarfs with $V_{rot}^{GF}$ lower than expected from the BTFR at their measured $\mbary$.}
\end{deluxetable}

\acknowledgments
We thank the anonymous referee for their detailed and useful feedback to help improve the original manuscript.\ KS acknowledges support from the Natural Sciences and Engineering Research Council of Canada (NSERC).

The Green Bank Observatory is a facility of the National Science Foundation operated under cooperative agreement by Associated Universities, Inc.

The Legacy Surveys consist of three individual and complementary projects: the Dark Energy Camera Legacy Survey (DECaLS; NOAO Proposal ID \# 2014B-0404; PIs: David Schlegel and Arjun Dey), the Beijing-Arizona Sky Survey (BASS; NOAO Proposal ID \# 2015A-0801; PIs: Zhou Xu and Xiaohui Fan), and the Mayall z-band Legacy Survey (MzLS; NOAO Proposal ID \# 2016A-0453; PI: Arjun Dey).\ DECaLS, BASS and MzLS together include data obtained, respectively, at the Blanco telescope, Cerro Tololo Inter-American Observatory, National Optical Astronomy Observatory (NOAO); the Bok telescope, Steward Observatory, University of Arizona; and the Mayall telescope, Kitt Peak National Observatory, NOAO.\ The Legacy Surveys project is honored to be permitted to conduct astronomical research on Iolkam Du'ag (Kitt Peak), a mountain with particular significance to the Tohono O'odham Nation.

NOAO is operated by the Association of Universities for Research in Astronomy (AURA) under a cooperative agreement with the National Science Foundation.

This project used data obtained with the Dark Energy Camera (DECam), which was constructed by the Dark Energy Survey (DES) collaboration.\ Funding for the DES Projects has been provided by the U.S.\ Department of Energy, the U.S.\ National Science Foundation, the Ministry of Science and Education of Spain, the Science and Technology Facilities Council of the United Kingdom, the Higher Education Funding Council for England, the National Center for Supercomputing Applications at the University of Illinois at Urbana-Champaign, the Kavli Institute of Cosmological Physics at the University of Chicago, Center for Cosmology and Astro-Particle Physics at the Ohio State University, the Mitchell Institute for Fundamental Physics and Astronomy at Texas A\&M University, Financiadora de Estudos e Projetos, Fundacao Carlos Chagas Filho de Amparo, Financiadora de Estudos e Projetos, Fundacao Carlos Chagas Filho de Amparo a Pesquisa do Estado do Rio de Janeiro, Conselho Nacional de Desenvolvimento Cientifico e Tecnologico and the Ministerio da Ciencia, Tecnologia e Inovacao, the Deutsche Forschungsgemeinschaft and the Collaborating Institutions in the Dark Energy Survey.\ The Collaborating Institutions are Argonne National Laboratory, the University of California at Santa Cruz, the University of Cambridge, Centro de Investigaciones Energeticas, Medioambientales y Tecnologicas-Madrid, the University of Chicago, University College London, the DES-Brazil Consortium, the University of Edinburgh, the Eidgenossische Technische Hochschule (ETH) Zurich, Fermi National Accelerator Laboratory, the University of Illinois at Urbana-Champaign, the Institut de Ciencies de l'Espai (IEEC/CSIC), the Institut de Fisica d'Altes Energies, Lawrence Berkeley National Laboratory, the Ludwig-Maximilians Universitat Munchen and the associated Excellence Cluster Universe, the University of Michigan, the National Optical Astronomy Observatory, the University of Nottingham, the Ohio State University, the University of Pennsylvania, the University of Portsmouth, SLAC National Accelerator Laboratory, Stanford University, the University of Sussex, and Texas A\&M University.

BASS is a key project of the Telescope Access Program (TAP), which has been funded by the National Astronomical Observatories of China, the Chinese Academy of Sciences (the Strategic Priority Research Program "The Emergence of Cosmological Structures" Grant \# XDB09000000), and the Special Fund for Astronomy from the Ministry of Finance.\ The BASS is also supported by the External Cooperation Program of Chinese Academy of Sciences (Grant \# 114A11KYSB20160057), and Chinese National Natural Science Foundation (Grant \# 11433005).

The Legacy Survey team makes use of data products from the Near-Earth Object Wide-field Infrared Survey Explorer (NEOWISE), which is a project of the Jet Propulsion Laboratory/California Institute of Technology.\ NEOWISE is funded by the National Aeronautics and Space Administration.

The Legacy Surveys imaging of the DESI footprint is supported by the Director, Office of Science, Office of High Energy Physics of the U.S.\ Department of Energy under Contract No.\ DE-AC02-05CH1123, by the National Energy Research Scientific Computing Center, a DOE Office of Science User Facility under the same contract; and by the U.S.\ National Science Foundation, Division of Astronomical Sciences under Contract No.\ AST-0950945 to NOAO.

\vspace{5mm}
\facilities{GBT (VEGAS)}

\software{astropy\citep{astropy:2013,astropy:2018}, 
    GALFIT \citep{Galfit},
    GBTIDL \citep{GBTIDL}, 
    SEP \citep{2016SEP,1996SourceExtractor},
    }

\bibliographystyle{aasjournal}
\bibliography{references}

\end{document}